\documentclass[structabstract]{aa}
\usepackage{graphicx}
\usepackage{natbib}
\bibpunct{(}{)}{;}{a}{}{,} % to follow the A&A style
\usepackage{txfonts}
\usepackage[usenames]{color}
\begin{document}

   \title{Modelling a high-mass red giant observed by CoRoT\thanks{The CoRoT space mission, launched on 2006 December 27, was developed and is operated by the CNES with participation of the Science Programs of ESA; ESA's RSSD, Austria, Belgium, Brazil, Germany and Spain.}}

   \author{
   	F. Baudin\inst{1}\and
   	C. Barban\inst{2}\and
        M.J. Goupil\inst{2}\and
        	  R. Samadi\inst{2}\and
	  Y. Lebreton\inst{3}\and	  
	  H. Bruntt\inst{2}\and\\
	 T. Morel\inst{4}\and
          L. Lef\`evre\inst{2}\and
	  E. Michel\inst{2}\and
	  B. Mosser\inst{2}\and
          F. Carrier\inst{5}\and
          J. De Ridder\inst{5}\and \\
          A. Hatzes\inst{6}\and
          S. Hekker\inst{7,8}\and
          T. Kallinger\inst{5}\and 
          M. Auvergne\inst{2}\and
          A. Baglin\inst{2}\and
          C. Catala\inst{2}
          }
          
         \institute{ Institut d'Astrophysique Spatiale, UMR8617, CNRS, Universit\'e Paris XI, B\^atiment 121, 91405 Orsay Cedex, France
         \and LESIA, UMR8109, CNRS, Universit\'e P. et M. Curie, Universit\'e D. Diderot, Observatoire de Paris, 92195 Meudon Cedex, France
   	\and GEPI, UMR8111, CNRS, Universit\'e D. Diderot, Observatoire de Paris, 92195 Meudon Cedex, France
	\and Institut d'Astrophysique et de G\'eophysique, Universit\'e de Li\`ege, All\'ee du 6 Aout, 4000 Li\`ege, Belgium
         \and Instituut voor Sterrenkunde, Katholieke Universiteit Leuven, Celestijnenlaan 200 B, 3001 Hervelee, Belgium
         \and Th{\"u}ringer Landessternwarte, D-07778 Tautenburg, Germany
         	\and School of Physics and Astronomy, University of Birmingham,
	 Edgbaston, Birmingham B15 2TT, UK
	 \and Astronomical Institute ``Anton Pannekoek'', University of Amsterdam, Science Park 904, 1098 XH Amsterdam, the Netherlands
             }

   \date{Received ; accepted}

% \abstract{}{}{}{}{} 
% 5 {} token are mandatory
 
  \abstract
  % context heading (optional)
  % {} leave it empty if necessary  
   {The advent of space-borne photometers such as CoRoT and \textit{Kepler} has opened up new fields in asteroseismology. This is especially true for red giants as only a few of these stars were known to oscillate with small amplitude, solar-like  oscillations before the launch of CoRoT.}
  % aims heading (mandatory)
   {The G6 giant HR\,2582 (HD\,50890) was observed by CoRoT for approximately 55 days. We present here the analysis of its light curve and the characterisation of the star using different observables, such as its location in the Hertzsprung-Russell diagram and seismic observables.}
  % methods heading (mandatory)
   {Mode frequencies are extracted from the observed Fourier spectrum of the light curve. Numerical stellar models are then computed to determine the characteristics of the star (mass, age, etc...) from the comparison with observational constraints.}
  % results heading (mandatory)
   {We provide evidence for the presence of solar-like oscillations at low frequency, between 10 and 20\,$\mu$Hz, with a regular spacing of $(1.7\pm0.1)\mu$Hz between consecutive radial orders. Only  radial modes are clearly visible.
From the models compatible with the observational constraints used here, 
We find that HR\,2582 (HD\,50890) is a massive star with a mass in the range (3--\,5\,$M_{\odot}$), clearly above the red clump. It oscillates with rather low radial order ($n$ = 5\,--\,12) modes.
Its evolutionary stage cannot be determined with precision: the star could be on the ascending red giant branch (hydrogen shell burning) with an age of approximately 155 Myr or in a later phase (helium burning). 
In order to obtain a reasonable helium amount, the metallicity of the star must be quite subsolar.
Our best models are obtained with a mixing length significantly smaller than that obtained for the Sun with the same physical description (except overshoot).
The amount of core overshoot during the main-sequence phase is found to be mild, of the order of 0.1\,$H_{\rm p}$.}
  % conclusions heading (optional), leave it empty if necessary 
   {HR\,2582 (HD\,50890) is  an interesting case as only a few massive stars can be observed due to their rapid evolution compared to less massive red giants.
HR\,2582 (HD\,50890) is also one of the few cases that can be used to validate the scaling relations for massive red giants stars and its sensitivity to the physics of the star.}

   \keywords{Stars: oscillations -- Stars: individual: HR\,2582 (HD\,50890), HD\,50890, HIP\,33243}

   \maketitle
%
%__________________________________________________________

\section{Introduction}
\label{sec:intro}

Solar-like oscillations (p modes excited by turbulent motion in the convective outer layers of the star) have first been observed in a couple of red-giant stars using either spectroscopic data from the ground \citep[see e.g.][]{Bedding07} or space photometric data \citep{Barban07}.
The detection of such oscillations in evolved stars, such as red giants, is made difficult by their very low frequencies (generally below 100\,$\mu$Hz), but made easier by their intrinsically larger amplitudes (around 100\,ppm in photometry) and smaller linewidths compared to solar-like stars. The use of long, uninterrupted observations provided by the space-based photometer CoRoT \citep{Baglin06} has permitted the detection of oscillations in hundreds of red giants \citep{deRidder09}. The observed spectra show different characteristics: regularly spaced frequencies or more complex spectra \citep{Hekker09}.
Analysing {\it Kepler} data, \citet{Bedding10} found many of these regularly spaced frequency spectra. Based on the same mission results, the fundamental seismic parameters of red giants were described by \citet{Kallinger10}.
Observed complex or regularly spaced spectra can be compared to the theoretical expectations of \citet{Dupret09} and can be understood as the presence or the absence of mixed modes. The regular spectra can be described using a ``universal pattern'' \citep{Mosser11}. The observed pulsating stars described in \citet{Hekker09} show a distribution of seismic parameters that indicate that they belong to the so-called ``red clump'' of low mass, He-burning, evolved stars \citep{Miglio09}.
The determination of the mass, radius and age of these evolved stars is a challenge \citep{Kallinger09} when their luminosity and effective temperature are poorly estimated. Scaling relations describing seismic characteristics are then used to determine their mass and radius.
Few individual red-giant stars have been modelled using seismic constraints. For example, \citet{Miglio10} found the signature of sharp structure variation in the red giant HR\,7349 observed by CoRoT and analysed by \citet{Carrier10}. \citet{diMauro11} also modelled a red giant observed by {\it Kepler} that was found in the hydrogen-shell burning phase.

   \begin{figure}
   \centering
\includegraphics[width=8cm]{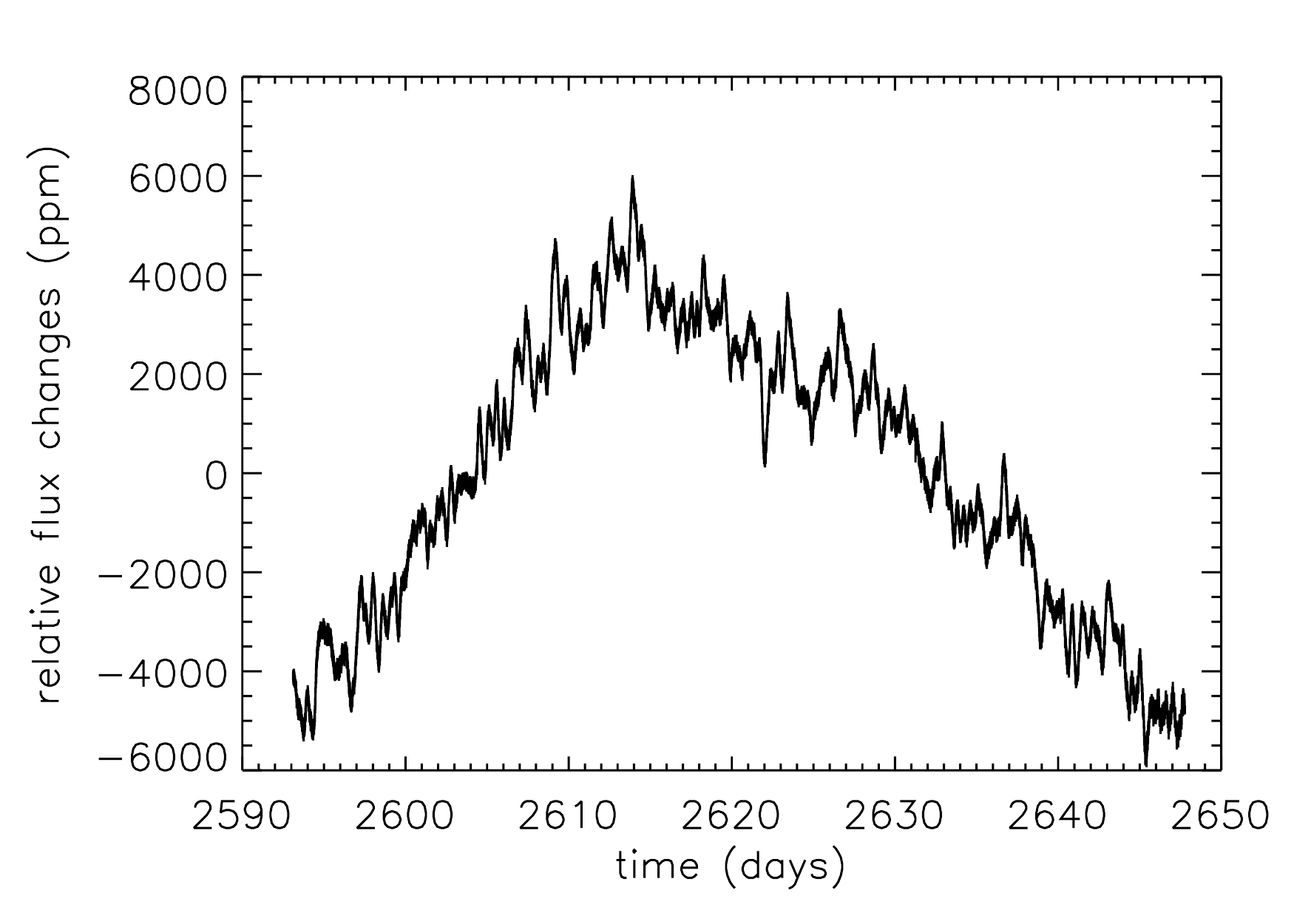}
      \caption{Time series of the relative intensity variations of HR\,2582 (HD\,50890) obtained by CoRoT.}
         \label{lc}
   \end{figure}

The hundreds of red giants described by \citet{Hekker09} were observed in the so-called ``exo-field'' of CoRoT (intended for a large number of faint stars) but CoRoT observed HR\,2582 (HD\,50890) in its seismo-field (used for a few bright stars). Observations lasted approximately $T_{\rm OBS}=55$\,days, during the first CoRoT observing run dedicated to science. Compared to the hundreds of red
giants observed in the CoRoT exo-field, HR\,2582 (HD\,50890) is an interesting case as its mass can also be inferred from non-seismic quantities. It therefore offers the opportunity of verifying the  compatibility with seismic scaling relations.

HR\,2582 (HD\,50890, HIP\,33243) is listed in the Simbad catalogue at CDS\footnote{Centre de Donn\'ees de Strasbourg,\\ \tt www.simbad.u-strasbg.fr/simbad/} as a G6.0\,III star with an apparent magnitude $m_V=6.03$. Some of its characteristics were derived from complementary observations detailed in Sect.\,\ref{sec:spectro}, which yielded: $T_{\rm eff}=4665\pm200$\,K, ${\rm [Fe/H]} = -0.18 \pm 0.14$, $\log L/L_{\odot}=2.70\pm 0.15$, and a radius of $R/R_{\odot}\sim 34 \pm 8$.
Based on the asteroseismic scaling relations (Eq.\,\ref{eq:numax},\,\ref{eq:largesep}), it appears that HR\,2582 (HD\,50890) is more massive than the stars belonging to the red clump mentioned above. \textbf{This makes this star a rare case since its high mass implies a rapid evolution. Modelling this star brings insights on the processes at work  in the last stages of evolution when the star could be burning its hydrogen in a shell or its central helium or even the later stage of shell-He burning. The brightness and parallax of HR\,2582 (HD\,50890) afford for relatively tight constraints on its position in the Hertzsprung-Russell (H-R) diagram (see Sect.\,\ref{sec:spectro}). This is not enough to determine if the star is ascending or descending the red giant branch (the determination of the observed evolutionary stage will require detailed seismic information, see Sect.\,\ref{sec:age}) but this nevertheless imposes in turn strong constraints on the modelling of the star's internal structure. This is also a good opportunity to assess the uncertainties on the results of such modelling.}

   \begin{figure}
   \centering
\includegraphics[width=8cm,angle=90]{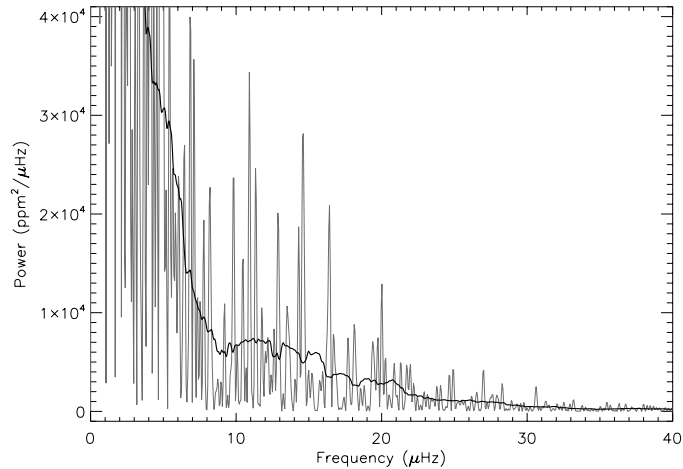}
      \caption{\textbf{Oversampled (by a factor of 4)} power density spectrum of the light curve shown in Fig.~\ref{lc} and a smoothed version in thick solid line (using a 1.8-$\mu$Hz boxcar).}
         \label{ps}
   \end{figure}

Our purpose here is to determine more precisely the characteristics of the star. We first use the mean large separation $\langle\Delta\nu\rangle$ and the frequency of maximum oscillation amplitude ($\nu_{\rm max}$) to 
determine the mass and radius of the star from scaling relations together with associated error bars. 
We study their compatibility within the error bars with
masses and radii derived from the location of the star in the H-R diagram. These last values  also 
suffer from uncertainties in the location in the H-R diagram. In addition, 
they also depend on the physical description used. Thus, uncertainties
in the physics must be accounted for in the uncertainties derived for the (non-seismic) mass and radius of the star.
Despite all of these uncertainties, we show that combining both seismic and non-seismic information results in a much more precise
determination of mass and radius than considering only one of these two types of  information. In addition, we are able to
provide a tighter constraint on the metallicity of the star. We also derive 
constraints on the values of the free parameters entering the 
description of convection adopted here.

Details of the observations and data analysis are given in Sect.\,\ref{sec:datanalysis}. The seismic spectrum 
is discussed in Sect.\,\ref{sec:spec}. A first approach to the modelling of the star is discussed in Sect.\,\ref{sec:model} 
where the influence of the physical description and parameters used in the models are estimated, before an optimal model is 
searched for in Sect.\,\ref{sec:model2}. Age ambiguity  for HR\,2582 (HD\,50890) is discussed in Sect.\,\ref{sec:age}. Conclusions are given in Sect.\,\ref{sec:conclu}.
 
   \begin{figure}
   \centering
\includegraphics[width=7cm,angle=90]{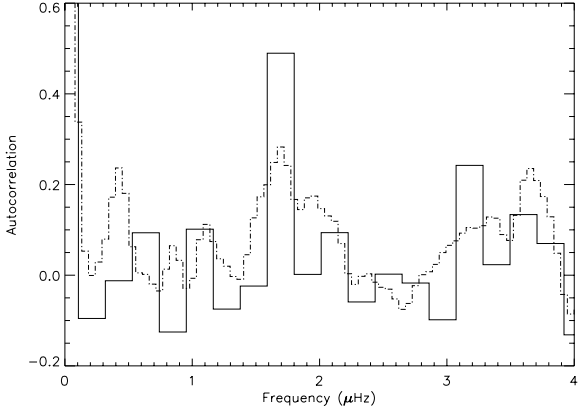}
      \caption{Autocorrelation of the power density spectrum between 9 and 24 $\mu$Hz in solid line \textbf{(and using an oversampled spectrum in dashed line)}}
         \label{autoc}
   \end{figure}

\section{Observations}

\subsection{CoRoT photometric data}
\label{sec:datanalysis}

A detailed description of the CoRoT instrument and its performance is given by \citet{Auvergne09}.

HR\,2582 (HD\,50890) has been observed for 54.65 days during the first observing run dedicated to science at the beginning of 2007.
We use in this paper the data ready for scientific analysis\footnote{The ``HELREG level 2'' data corresponding to light curves that have been corrected for known instrumental effects  and resampled onto a regular cadence in the heliocentric frame with one measurement every 32 seconds \citep{Samadi07}}. Outliers and other missing data points were replaced by interpolated points. The influence of interpolation was checked to be negligible. The duty cycle before interpolation was 90\%. The light curve is shown in Fig.~\ref{lc}. We notice in Fig.~\ref{lc} a clear inverse V-shape that is not explained by any known instrumental effect and that is not seen in the light curves obtained for other stars during the same observing run. This inverse V-shape, whatever is its origin, is a long-term trend that can be removed without consequence to our analysis since we are interested in periods of less than 1 day.
Thus, the data are high-pass filtered by substracting two different straight lines, one for the left part and one for the right part of the V-shape.

The power density spectrum of the time series, computed using a Fast Fourier Transform, is shown in Fig.\,\ref{ps}. \textbf{After oversampling by a factor of 4,}
it has a frequency resolution of $0.05\mu$Hz and it is normalised such that the total power integrated from zero to twice the Nyquist frequency is equal to the variance of the residual light curve; or in other words, the Fourier transform is normalised by $1/\sqrt{T}$ where $T$ is the total length of the observations.
%The power density spectrum of the detrended light curve is shown in Fig.\,\ref{ps}.
   \begin{figure}
   \centering
\includegraphics[width=7cm,angle=90]{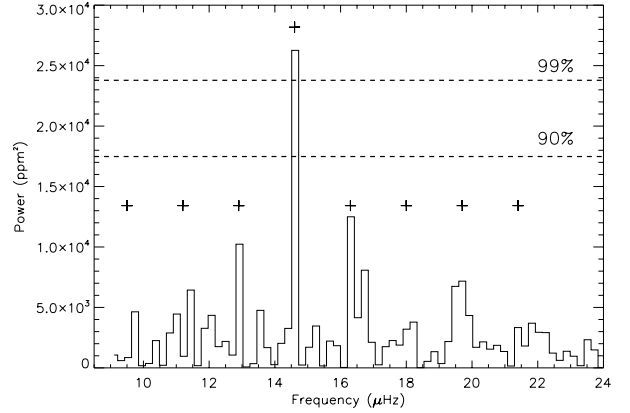}
      \caption{Power spectrum of the light curve shown in Fig.~\ref{lc}, once the contribution of the background noise has been removed (see text). \textbf{No oversampling is used in order to compute the statistical levels of confidence.}
The dashed lines indicate the 99\% and 90\% levels of confidence for detection (see text).
The crosses indicate frequencies that are equally spaced by the value found from the power autocorrelation, 1.7 $\mu$Hz.
%Default sampling ($1/T_{\rm OBS}$) is used because oversampling does not bring any improvement.
}
         \label{psz}
   \end{figure}
   
\subsection{Complementary observations}
\label{sec:spectro}

The Hipparcos parallax of HR\,2582 (HD\,50890) is $\pi=2.99\pm 0.44$\,mas according to the new determination of \citet{vanLeeuwen07}.

To determine the atmospheric parameters of HR\,2582 (HD\,50890), we analysed a spectrum from CORALIE \citep{Queloz00} obtained on 20 November 2001, which is available through the GAUDI database \citep{Solano05}. The spectrum has a resolution of 45000 and signal-to-noise ratio in the continuum around 6000\,\AA\ of 160. The lines are relatively broad and we measure $v \sin i = 10\pm2$ km/s. The broad lines made it difficult to do the analysis due to blending, but our results are based on the calculation of synthetic spectra that take the blending into account: we used the VWA software \citep{Bruntt10a,Bruntt10b}.

We used 44 Fe\,\textsc{i} (neutral iron) lines while only 2 Fe\,\textsc{ii} lines were available. To constrain $\log g$ one often uses the condition that Fe\,\textsc{i} and Fe\,\textsc{ii} abundances must agree. However, this was not possible since so few lines of Fe\,\textsc{ii} were available.
However, this approach was not very robust since only three partially blended Fe\,\textsc{ii} were available (6084.1, 6247.6, 6517.0\,\AA). Also, the wide Ca lines at 6122 and 6162\,\AA\, could not be used to further constrain $\log g$, due to their lack of sensitivity.
%We determined Teff, logg and [Fe/H] to be 4665+-200 K, $\log g = 1.4\pm0.3$, and [Fe=H] = -0.18 +-0.14.
%In addition, the wide Ca lines at 6122 and 6162\,\AA\ could not be used to constrain $\log g$, due to their lack of sensitivity.
With this lack of constraints, we found $T_{\rm eff}=4665\pm200$, $\log g = 1.4\pm0.3$ and ${\rm [Fe/H]} = -0.18 \pm 0.14$.

On the basis of these data, we estimated a bolometric correction of $-0.40\pm0.12$ mag following \citet{Vandenberg03}. It yields a luminosity of $\log L/L_{\odot}=2.70\pm 0.15$ and, through the Stefan-Boltzmann law, a radius of $R/R_{\odot}\sim 34 \pm 8$.

\section{Interpretation of the seismic spectrum}
\label{sec:spec}

In the power density spectrum, an excess of power is clearly seen around $\nu_{\rm max}=(15 \pm 1)\mu$Hz, showing a bell-shape, enhanced in the smoothed version of the power spectrum, a well-known property of a solar-like oscillation spectrum. The central frequency $\nu_{\rm max}$ of this observed excess of power can be used to determine a first estimate of the mass of the star, based on the scaling relation suggested by \citet{Brown91} and derived by \citet{kb05}: 

%   \begin{equation}
%  R_* = \frac{\left( \nu_{\rm max}/\nu_{\rm max, \odot} \right) \sqrt{T_{\rm eff}/T_{\rm eff \odot}}}{\left(\Delta \nu/\Delta \nu_{\odot}\right)^2}
%   \end{equation}
%   \begin{equation}
%  M_* = \frac{\left( \nu_{\rm max}/\nu_{\rm max, \odot} \right)^3 \left(T_{\rm eff}/T_{\rm eff \odot}\right)^{3/2}}{\left(\Delta \nu/\Delta \nu_{\odot}\right)^4}
%   \end{equation}

      \begin{equation}
   \frac{\nu_{\rm max}}{\nu_{\rm max, \odot}}  \approx \frac{M/M_{\odot}}{(R/R_{\odot})^2 \sqrt{T_{\rm eff}/T_{\rm eff \odot}}} 
   \label{eq:numax}
   \end{equation}

From the stellar parameters of HR\,2582 (HD\,50890) given in Sect.\,\ref{sec:intro}, we derive a first estimate of the mass of approximately $5.2 \pm 2.9 M_{\odot}$.\\
We then derive the statistical significance of the peaks observed in the power spectrum \textbf{(without oversampling, thus giving a resolution of $0.21\mu$Hz)}, using the so-called null hypothesis \citep{Fisher25}. The spectrum is considered to be made of white noise and a confidence level is computed from the probability of a peak caused by noise to reach a given level of power. To apply this, the contribution of the background noise in the power density spectrum between 9 and 24 $\mu$Hz was removed by fitting a power law of the frequency and dividing the observed raw spectrum by this. The resulting power density spectrum is shown in Fig.~\ref{psz}. The noise level was determined from the mean power in the  frequency range considered. Then, the significance threshold corresponding to a probability of 90 and 99\% for the peaks not to be caused by white noise was computed and are represented also in Fig.~\ref{psz}.
It appears that the highest peak of the spectrum, around 15 $\mu$Hz, is the only one above the 99\% level of confidence.

Another expected characteristic of a solar-like oscillation spectrum is its comb-like structure.
Thus, we searched for regularly spaced frequencies in the power spectrum by computing the autocorrelation of this spectrum (Fig.~\ref{autoc}). This autocorrelation clearly shows an unambiguous signature of regularly spaced peaks with a mean spacing value of $\langle\Delta \nu\rangle\,=1.7\mu$Hz. This observed value can be compared with the expected value derived from the well known scaling relation and the first mass estimate from Eq.\,\ref{eq:numax}:
\begin{equation}
 \frac{\langle\Delta \nu\rangle}{\langle\Delta \nu_{\odot}\rangle}  \approx  \left(\frac{M}{M_{\odot}}\right)^{1/2} \left(\frac{R}{R_{\odot}}\right)^{-3/2} \; \Rightarrow \; \langle\Delta \nu\rangle \; \sim (1.7 \pm 0.4) \; \mu{\rm Hz}
\label{eq:largesep}
\end{equation}
which is in very good agreement with what is observed.

Then, coming back to the power spectrum, one can see that regularly spaced peaks follow the 1.7 $\mu$Hz spacing (indicated in Fig.~\ref{psz}). However, they have a low S/N ratio, between 1.2 and 2.3. Thus, from the simple null hypothesis, they cannot be considered as being modes, but the null hypothesis is designed to evaluate the confidence of detection of peaks only based on their S/N ratio. In the present case, the regular pattern that these peaks show is a strong argument to consider them as possible p modes. However, we considered that their frequencies were not very reliable because of this very low S/N ratio. Thus, 
the frequencies were extracted simply from the peak maxima in the power spectrum and are given in Table~\ref{tb1} and we considered the spectrum resolution ($\delta\nu=0.2\mu$Hz) to be the uncertainty of the frequencies. The individual frequencies will not be used for modelling of the star, which will use the mean large frequency separation instead.
From the frequency list, we can recompute the mean large frequency  separation and its standard deviation: $\langle\Delta \nu\rangle\,=(1.7 \pm 0.1)\mu$Hz.

\begin{figure}[t]
\centering
\includegraphics[trim= 100 150 100 100,width=6cm]{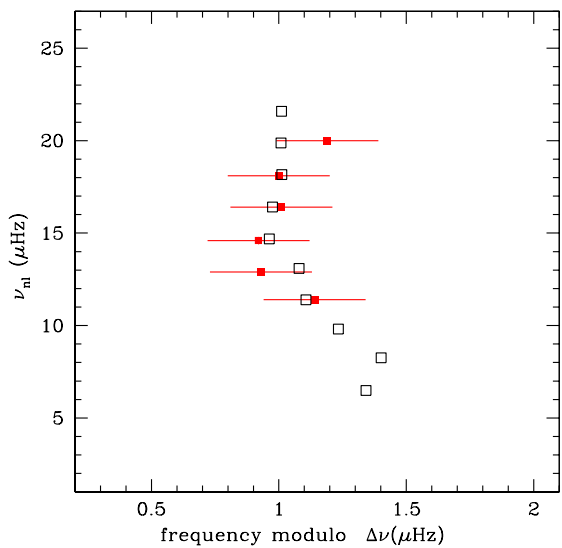}
\caption{Echelle diagram of the observed modes (in red) with $\Delta\nu=1.70\mu$Hz and the modes computed from the optimal model (in black)  with $\Delta\nu=1.72 \mu$Hz (see Sect.\,\ref{sec:diagech}).}
\label{fig:diagech}
\end{figure}

\begin{table}[h]
\caption{Frequency %and height
of the p modes identified using their regular frequency spacing.}
\begin{center}
%\begin{tabular}{c|c}
%Frequency ($\mu$Hz) & Height (ppm$^2$) \\
%\hline \hline
%%	9.73 $\pm$ 0.08  &     \\
%%	11.43 $\pm$ 0.04  &   \\
%%	12.91 $\pm$ 0.04  &    \\
%%	14.59 $\pm$ 0.03  &    \\
%%	16.35 $\pm$ 0.03  &    \\
%%	18.14 $\pm$ 0.08   &    \\
%%	19.58 $\pm$ 0.05   &    \\
%9.8	&   4635  \\
%11.4	&   6435  \\
%12.9	&   10230  \\
%14.6	&   26265  \\
%16.4	&   12495  \\
%18.2	&   3200  \\
%19.8	&   4330  \\ 
%21.9  &   3695  \\
\begin{tabular}{c}
\hline
Frequency \\
($\mu$Hz) \\
\hline
9.8	  \\
11.4	  \\
12.9	  \\
14.6	  \\
16.4	  \\
18.2	  \\
19.8	  \\ 
\hline
\end{tabular}
\caption{The uncertainty in frequency is conservatively considered to be the frequency resolution ($\delta\nu=0.2\mu$Hz).}
\label{tb1}
\end{center}
\end{table}

This result is obtained assuming that the observed peaks correspond to modes with the same degree $\ell$. For reasons of mode visibility, the most probable degrees to be detected are $\ell=0$ and $\ell=1$. It seems improbable that $\ell=0$ are not visible if $\ell=1$ modes are. However, in the present case, assuming the presence of $\ell=1$ modes would give a frequency separation twice as large and thus not agreeing with Eq.\,\ref{eq:largesep}.
Hence, we deduce that non-radial modes are not visible in the present observations of HR\,2582 (HD\,50890). \citet{deRidder09} and \citet{Bedding10} have shown that red giants generally show non-radial modes in seismic spectra similar to that of the Sun \citep[as also expected by][for stars of $2\,M_{\odot}$]{Dupret09}. In the present case, the relative shortness of the observations (55 days) may be at the origin of a too low resolution to detect non-radial modes (in particular mixed $\ell$=1 modes), or the relatively high mass of HR\,2582 (HD\,50890) leads to small amplitudes for non-radial modes. The low visibility of $\ell=1$ modes for some red giants is confirmed by {\it Kepler} observations \citep[see][]{Mosser11c}.

In addition, as observed by \citet{Hekker09} and discussed by \citet{Stello09}, the frequency spacing $\Delta \nu$ and the frequency of the maximum of the spectrum, $\nu_{\rm max}$, are linked. The comparison with  Fig.\,6 of \citet{Hekker09} shows that the observed value of $\nu_{\rm max}$ is in agreement with the frequency spacing observed if one considers that only radial modes are visible. More precisely, the $\Delta \nu$ observed for HR\,2582 (HD\,50890) corresponds to the lower values computed from the correlation observed by \citet{Hekker09}. This is another indication of the high mass of HR\,2582 (HD\,50890) because the relation between $\nu_{\rm max}$ and $\Delta \nu$ is sensitive to the mass \citep{Hekker11}.

By comparing $\nu_{\rm max}$ and $\Delta \nu$, one can estimate that this star oscillates with low-order $n$ modes (approximately $5 \leq n \leq 12$), which will be confirmed by modelling in Sect.\,\ref{sec:diagech}.

We then used the mean value of the large spacing to cut the power spectrum into pieces of $1.7 \mu$Hz. These pieces are then stacked on top of each other to build an \'echelle diagram, shown in Fig.~\ref{fig:diagech}, where modes with same degree $\ell$ appear in a ridge more or less vertically aligned. However, this ridge has a typical curvature that is observed in other stars, in particular in the Sun or solar-like pulsators and that will be compared with expectations drawn from stellar modelling in the following section.

\begin{figure*}
\centering
\includegraphics[width=8cm]{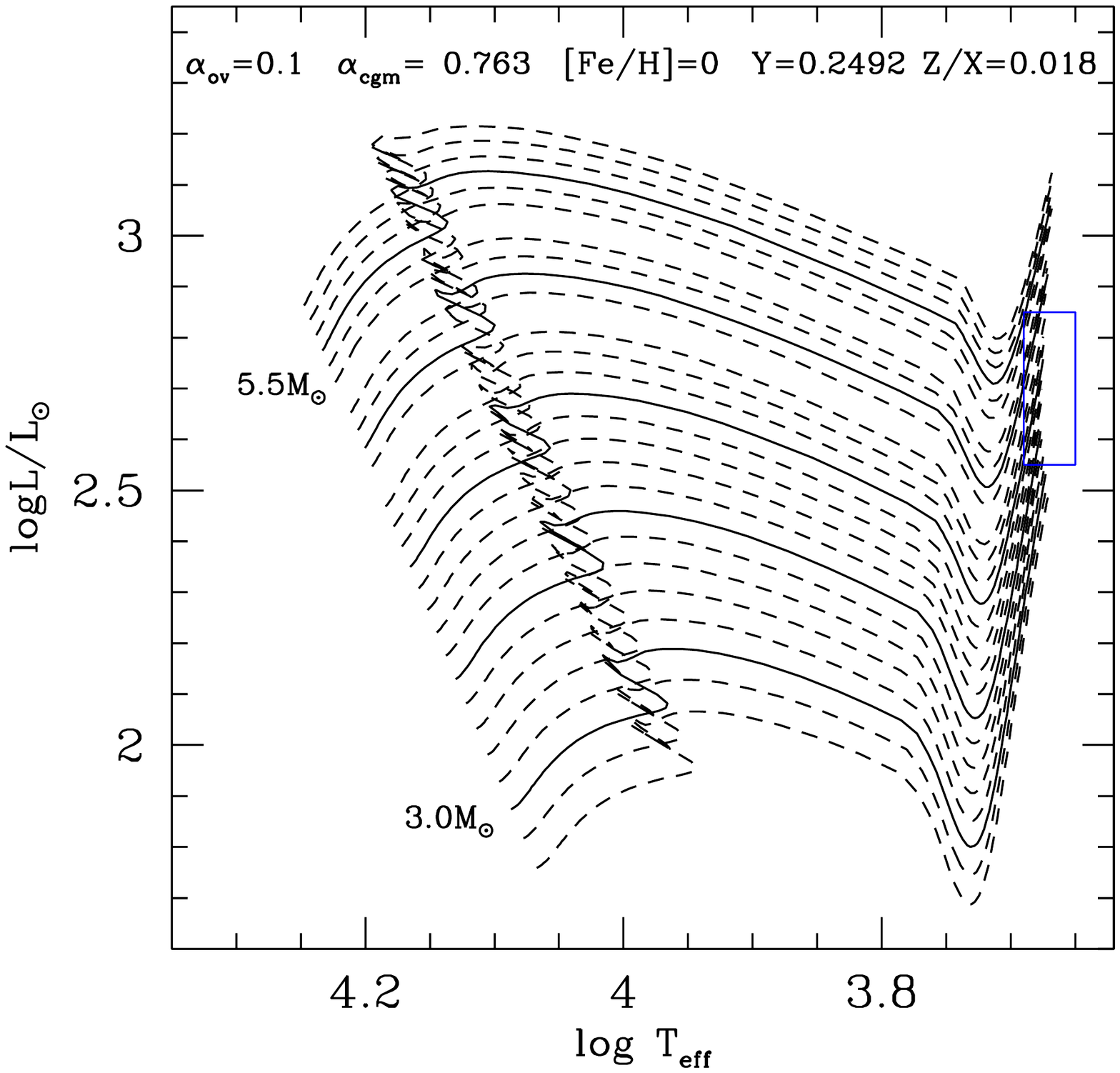}
\includegraphics[width=8cm]{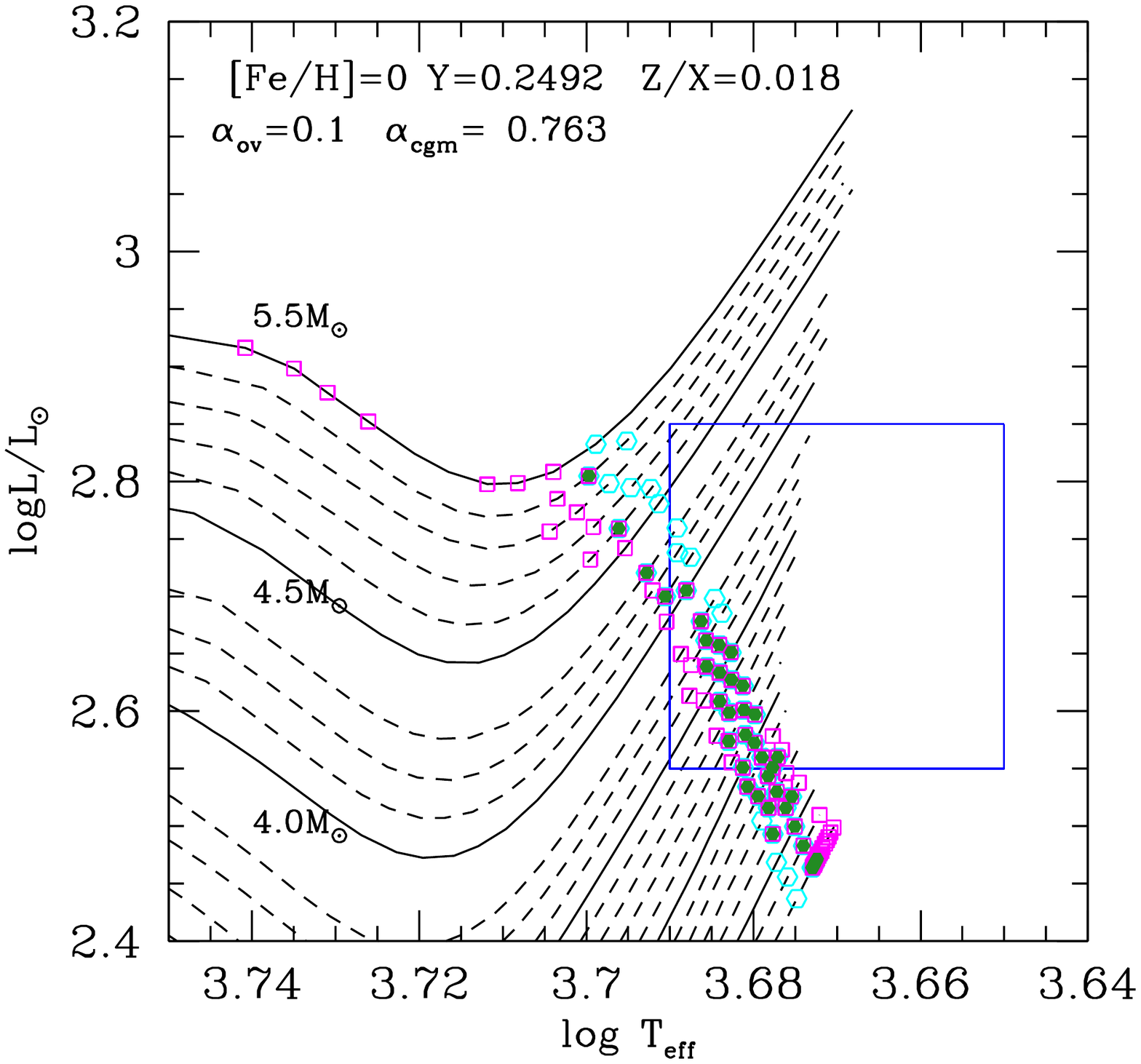}
\caption{Left: Evolutionary tracks for models with masses from 2.8 to 5.5\,$M_{\odot}$ and observational 
luminosity and effective temperature constraints (blue box);
Right: zoom of left panel with models satisfying the seismic constraints  shown with
 coloured points (cyan indicates models having a compatible $\nu_{\rm max}$, magenta a compatible $\Delta\nu$, and  dark green models satisfying both conditions). }
\label{fig:HR}
\end{figure*}

\section{Stellar characteristics for HR\,2582 (HD\,50890)}
\label{sec:model}

The modelling of the star is carried out using CESAM2k \citep{Morel97,Morel08}. In order to
compute a stellar model for a given mass and age (or luminosity and $T_{\rm eff}$), one needs to specify its initial relative helium abundance $Y$ and its metallicity $Z/X$. One must also provide values for the free parameters used in the physical description of the stellar interior.
In the present work, convection is included using the Canuto, Goldman, Mazzitelli (CGM) formulation \citep{Canuto96}.
This formulation requires a prescription for the characteristic scale length associated with the energy-bearing eddies.
Following \citet{Bernkopf98}, we assume that this  scale length is $\alpha_{\rm CGM} \, H_{\rm p}$ where $H_{\rm p}$ is the pressure scale height and $\alpha_{\rm CGM}$ a free parameter.

\begin{figure*}
\centering
\includegraphics[trim= 100 150 0 100,width=5.5cm]{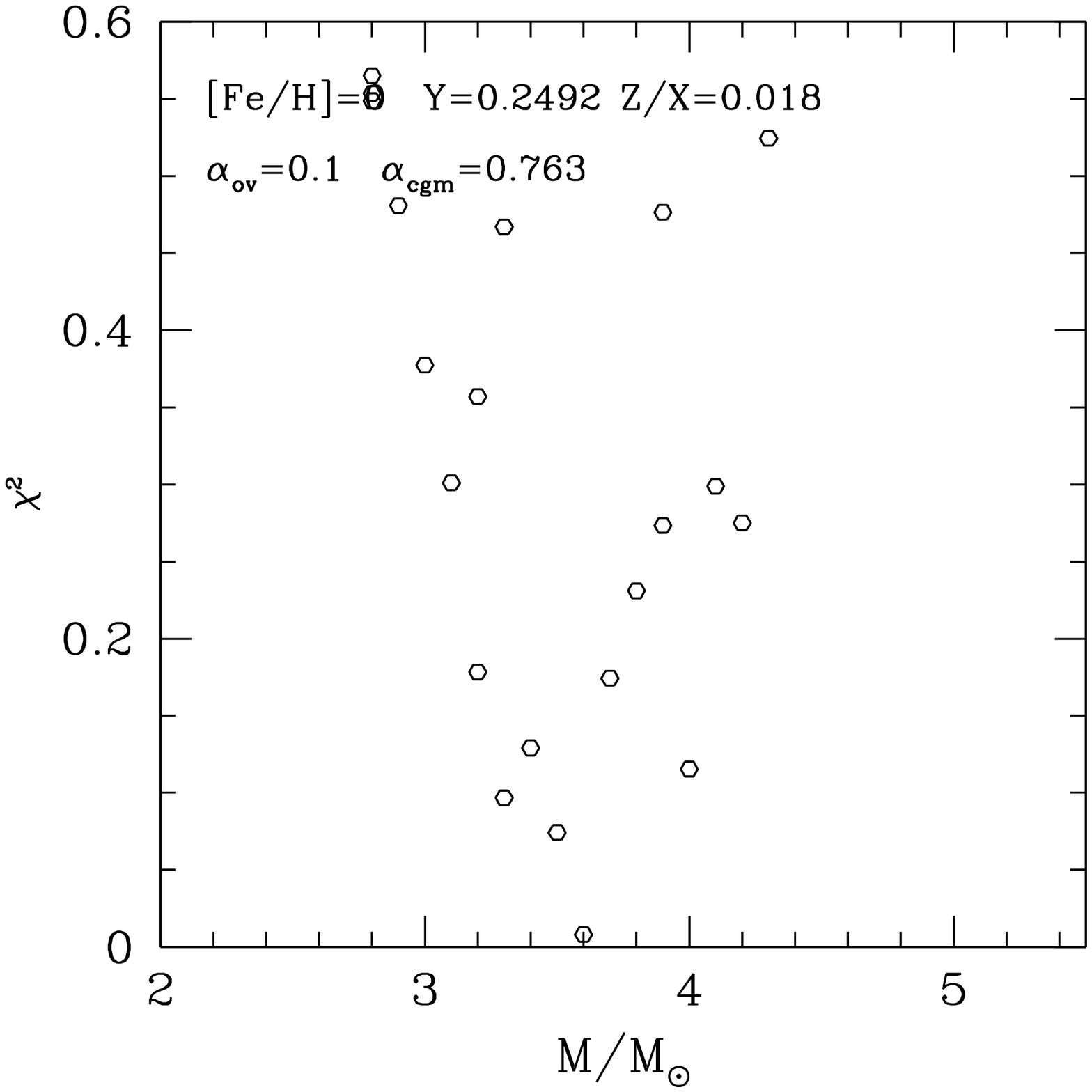}
\includegraphics[trim= 50 150 50 100,width=5.5cm]{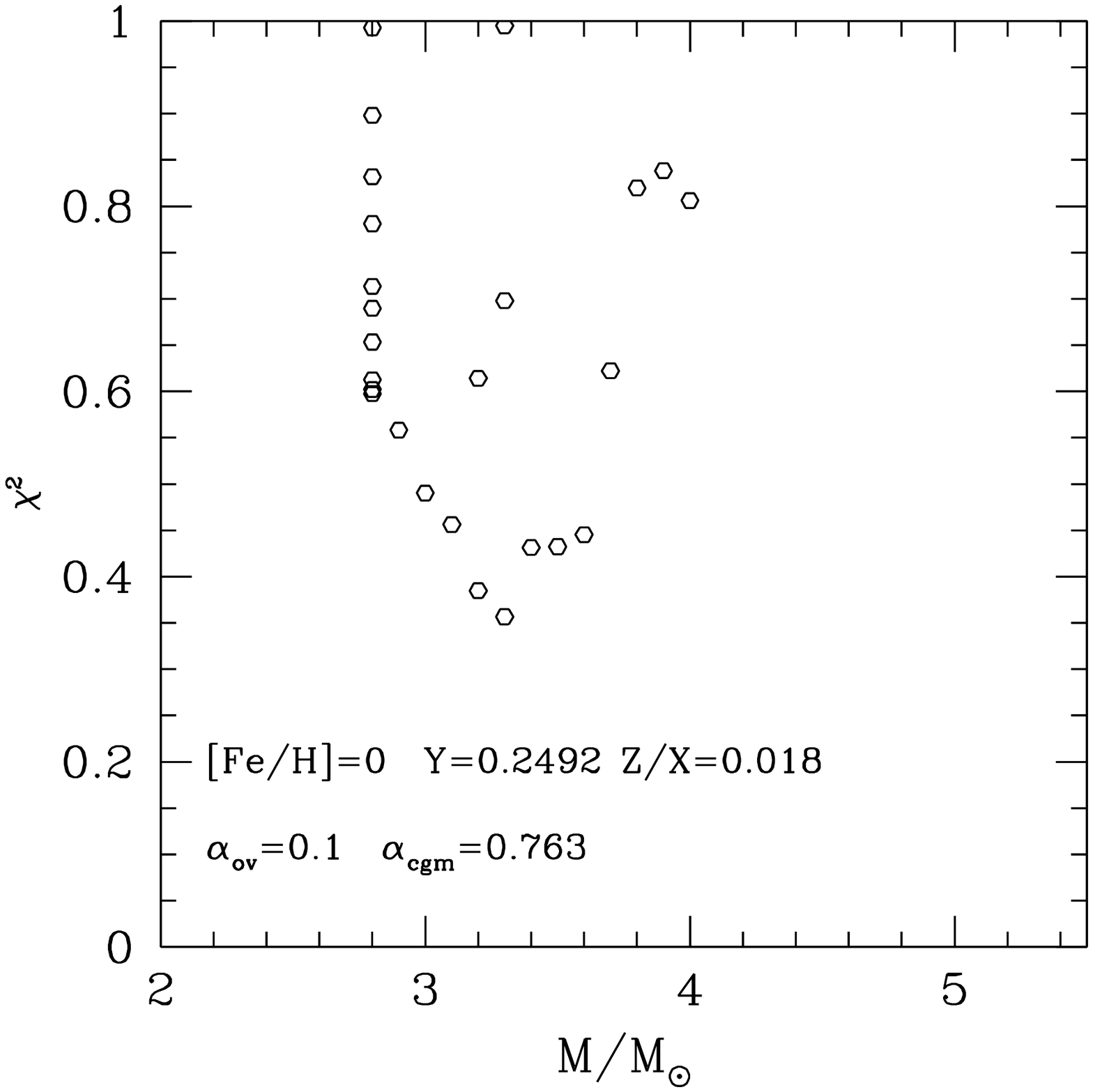}
\includegraphics[trim= 0 150 100 100,width=5.5cm]{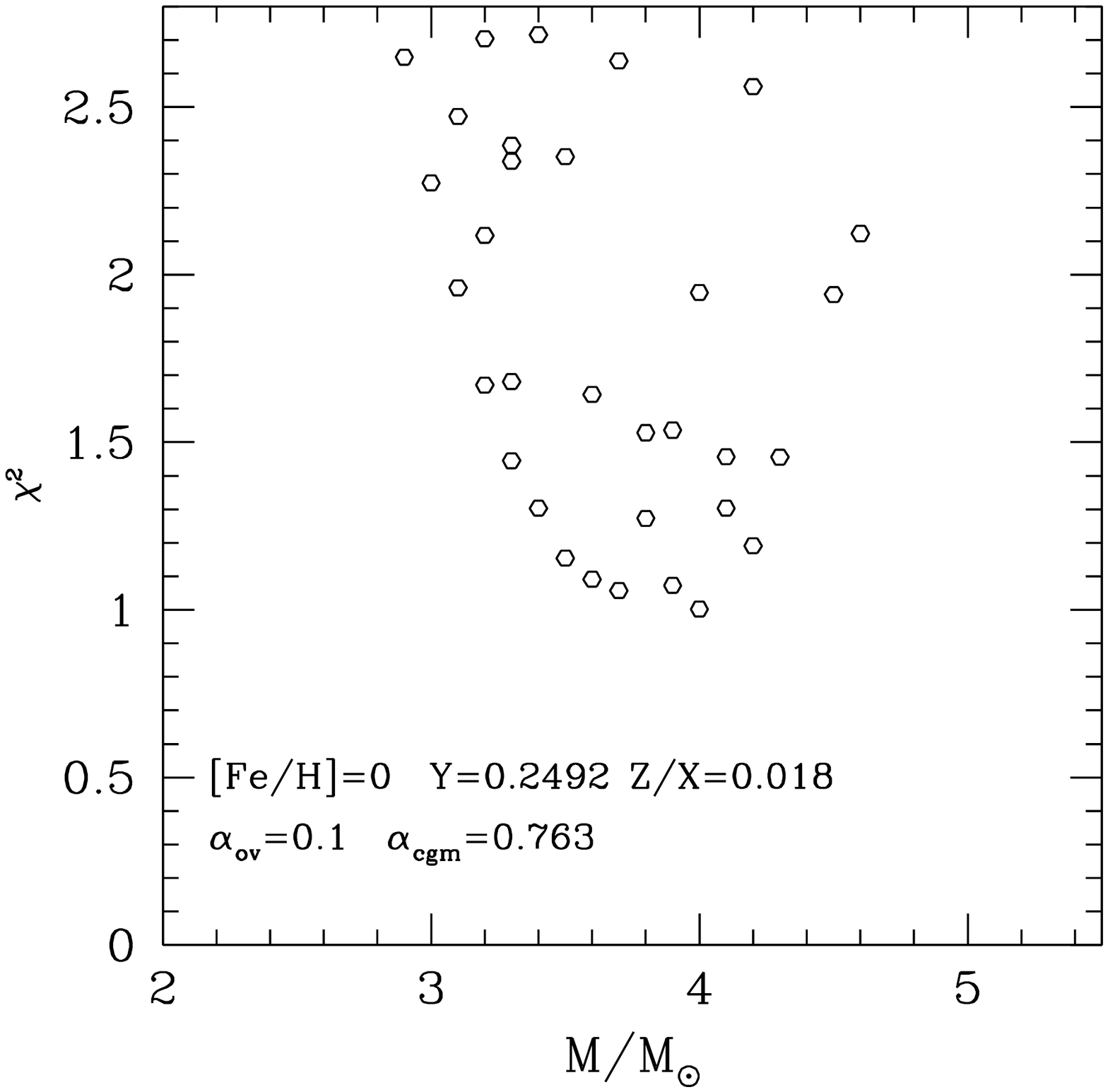}
\caption{Computed values of $\chi^2$ (Eq.\ref{eq:chi2}) for models corresponding to a given physical description as a function of the mass. 
On the left, the constraints are $\langle\Delta\nu\rangle>$ and $\nu_{\rm max}$,
 in the middle panel $T_{\rm eff}$ is added, and on the right 
 the luminosity is also taken into account. Note the 
 different scale in the latter case.}
\label{fig:opt}
\end{figure*}

Another free parameter is needed to describe convective overshoot since stars as massive as HR\,2582 (HD\,50890) have a convective core while on the main sequence.
The modification of the luminosity when convective core overshoot is included during the main sequence is maintained once core convection has disappeared after the main sequence.
Hence  core overshoot must be taken into account in modelling the red giant HR\,2582 (HD\,50890). Overshooting is described as an extension of $r_{\rm c}$, 
the radius of the core as defined according to the instability criterion of Schwarzschild, over a distance taken to be $d_{\rm ov} = \alpha_{\rm ov}\,{\rm min}(r_{\rm c}, H_{\rm p})$ where $\alpha_{\rm ov}$, the overshooting parameter, is our second free parameter.

The models considered are computed with the OPAL\,2005 equation of state and opacity tables as described in \citet{Lebreton08}.
The chemical mixture by \citet{Asplund05} is adopted for the calculation of the opacities.
For the calculation of the equation of state, we have assumed the chemical mixture of \citet{Grevesse98} with the exception of CNO for which values close to the recommendation of \citet{Asplund05} are adopted.
Neither microscopic diffusion nor near-surface effects are included in the modelling.

First, the star is modeled for a given physical description (fixed values for $\alpha_{\rm ov}$, $\alpha_{\rm CGM}$) and a given chemical composition. 
The ``best'' model is searched in a pre-existing grid of models for this description, 
defined as the model that  minimises the discrepancies with observational constraints, given this description (see Sect.\,\ref{sec:model1}). Then, the influence of  the chemical composition and physics parameters on the mass and radius of the best model are investigated individually (Sect.\,\ref{sec:model2}). Finally in Sect.\,\ref{sec:model2}, the minimisation of discrepancies 
between models and observation is performed by adjusting several of the free parameters.

\subsection{Modelling for a given physical description}
\label{sec:model1}

The models considered here are built  assuming the same chemical composition as the Sun, that is the metallicity $Z/X=0.018$ ($Z=0.0135$) and helium abundance  $Y = 0.2492$ and the same 
mixture. We also assume $\alpha_{\rm CGM}=0.763$, which is the value obtained for the Sun with the physics used here. 
As HR\,2582 (HD\,50890) is expected to have a convective core, we further assume convective core overshoot with $\alpha_{\rm ov}=0.1$ on the main sequence.
Fig.\,\ref{fig:HR} shows that stars with different masses (from $\sim 2$ to $\sim 5.5 M_{\odot}$)
have  evolutionary tracks compatible with the observational contraints $L$ and $T_{\rm eff}$. 
In this range, higher masses correspond to less evolved stars. All of the models describe stars on the ascending red-giant branch. However, models describing more evolved stars, at the helium-burning stage, can have the same position in the H--R diagram.
This point will be discussed in Sect.\,\ref{sec:age}.

\begin{figure*}
\centering
\includegraphics[trim= 100 150 0 100,width=5.5cm,angle=0]{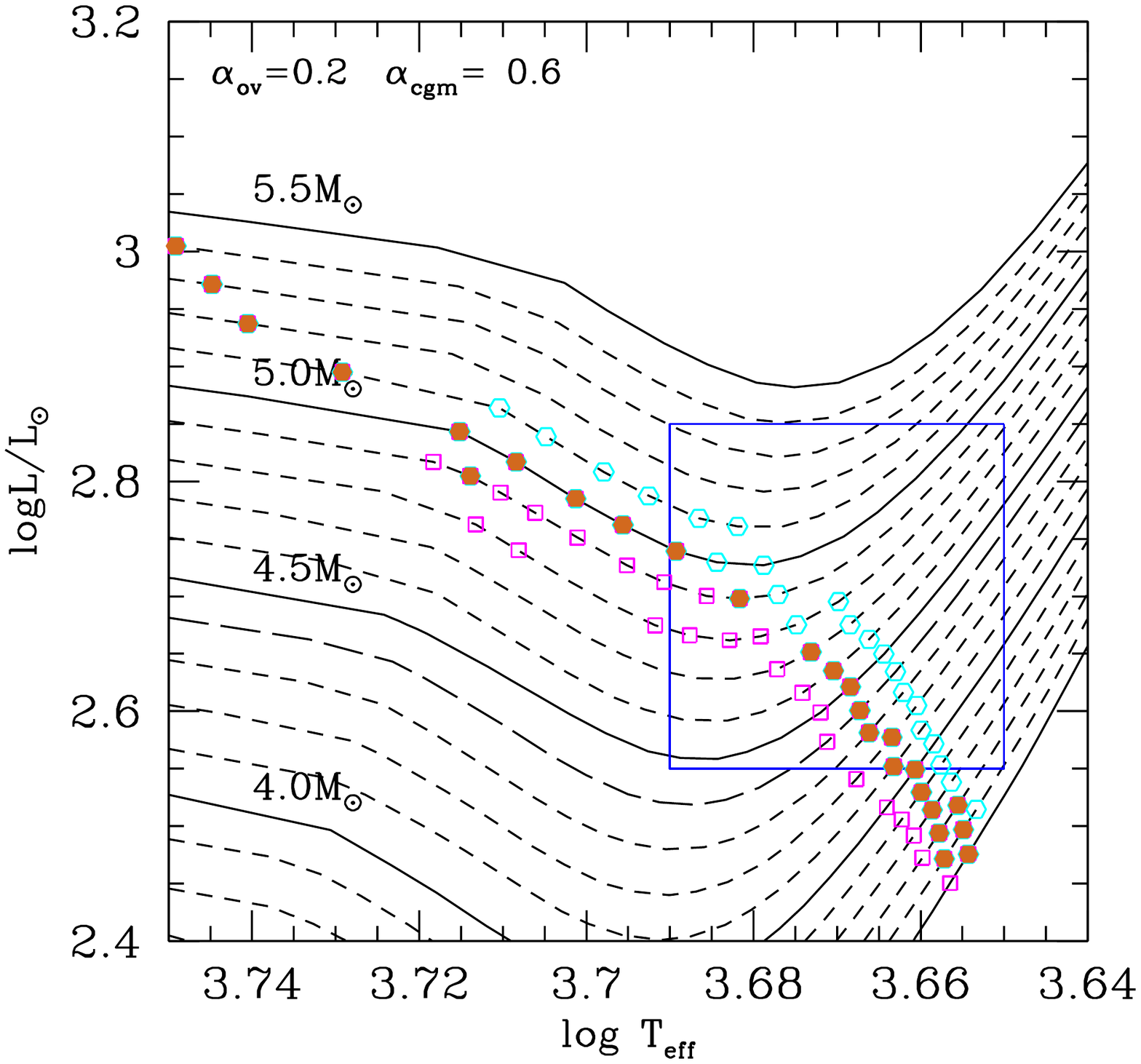}
\includegraphics[trim= 50 150 50 100,width=5.5cm,angle=0]{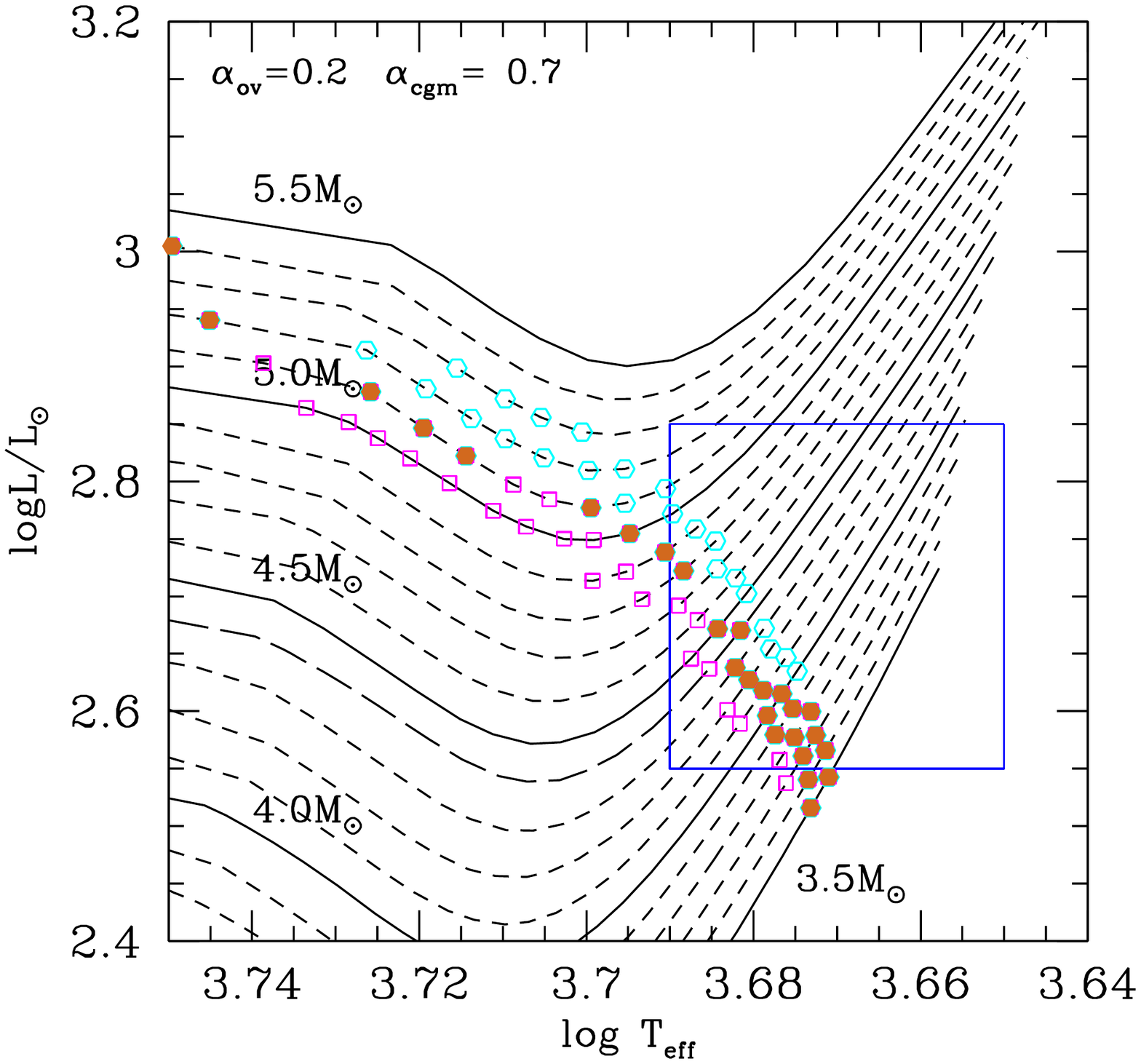}
\includegraphics[trim= 0 150 100 100,width=5.5cm,angle=0]{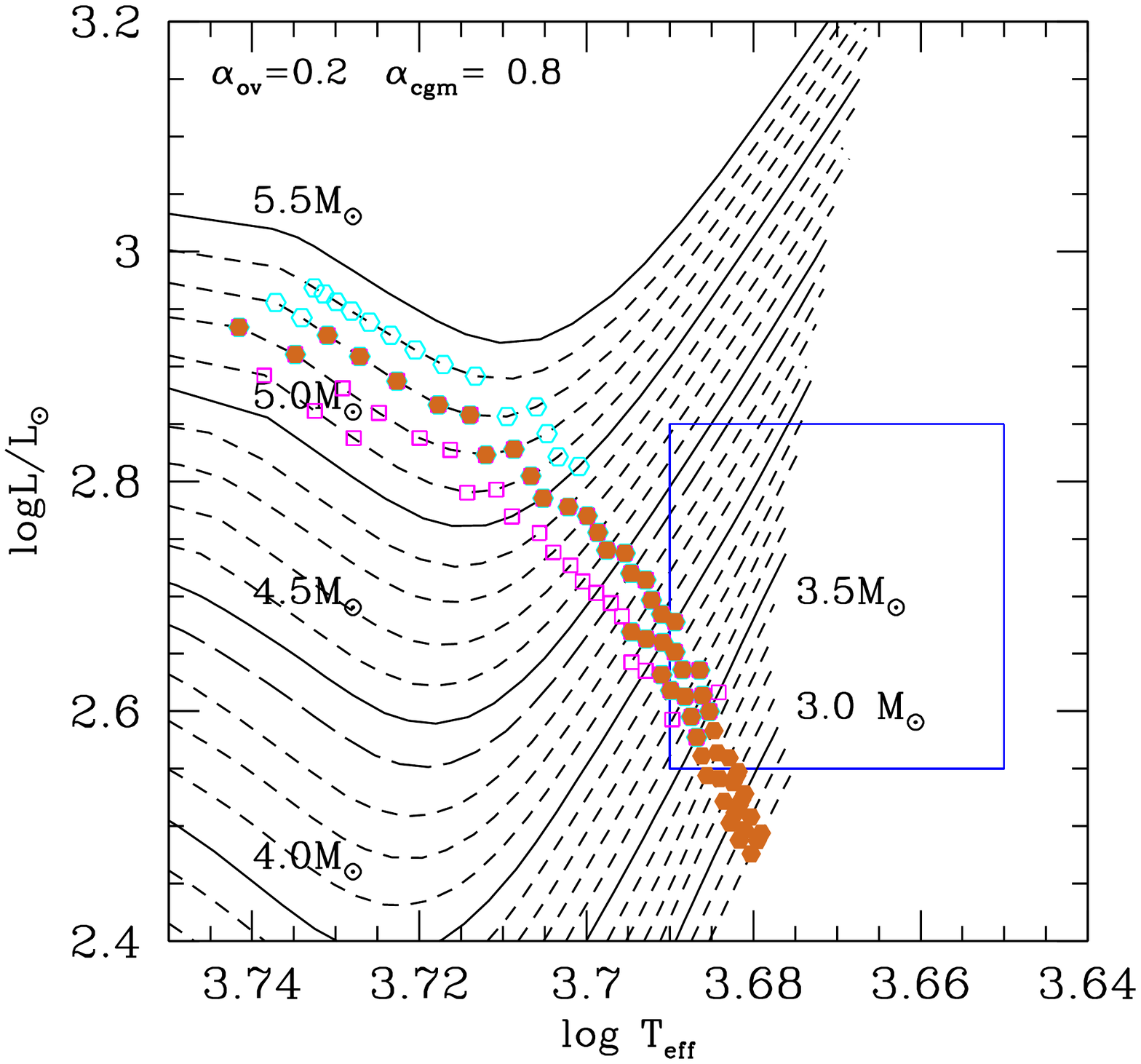}
\caption{Evolutionary tracks for different masses and different $\alpha_
{\rm CGM}$ (left: 0.6; center: 0.7; right: 0.8). H-R constraints ($L$ and $T_{\rm eff}$) are shown with the blue box, and models satisfying the seismic constraints are shown with coloured points (cyan indicates models having a compatible $\nu_{\rm max}$, magenta a compatible $\langle\Delta\nu\rangle$, and orange models satisfying both conditions). Models with higher values of $\alpha_{\rm CGM}$ shift compatible masses to lower values (see text).}
\label{fig:acgm}
\end{figure*}

Seismic information, such as the mean large separation, 
allows a more precise characterisation. The mean large separation links the mass to the radius through Eq.\,\ref{eq:largesep} and thus corresponds to a given strip in the H--R diagram (see for example Fig.\,\ref{fig:acgm}).
This allows a strong limitation of the possible observational values,
which is quite useful as the luminosity is poorly estimated because of the 
relatively large uncertainty on the parallax of HR\,2582 (HD\,50890).
We compute the mean large separation $\langle\Delta \nu\rangle$ and $\nu_{\rm max}$ according to Eq.\,\ref{eq:numax} and \ref{eq:largesep} for our model grids. The acceptable models (that is those that  satisfy  the observational constraint on $\langle\Delta \nu\rangle$) correspond to the highest temperatures and lowest luminosities of the observational range  for HR\,2582 (HD\,50890) and lead to a better mass estimate: $\sim$\,3 to $\sim$\,4\,$M_{\odot}$.
This is confirmed by the additional constraint provided by the frequency of 
the maximum power, $\nu_{\rm max}$. This might not come as a surprise as these two observables have been shown 
to be correlated for cool giants \citep{Hekker09,Stello09}.

We now wish to determine which model provides 
the ``best'' match with all of the observations. For all of the grid  models, we compute the quantity $\chi^2$:
\begin{equation}
\chi^2 = \sum_j \left( \frac{X^{\rm obs}_j - X^{\rm mod}_j}{\sigma_j} \right)^2
\label{eq:chi2}
\end{equation}
where $X_j$ are observables ($obs$ and $mod$ specifying the observed or the modelled value) and $\sigma_j$ the associated uncertainty.
Computations of $\chi^2$ were first performed with $X_1=\,\langle\Delta\nu\rangle$ and $X_2=\nu_{\rm max}$,
%(derived from Eq.\,\ref{eq:numax} and \ref{eq:largesep})
then adding $X_3=T_{\rm eff}$ and finally also taking into account $X_4=\log L/L_\odot$. The theoretical seismic quantities $\langle\Delta\nu\rangle$ and $\nu_{\rm max}$ are computed using Eq.\,\ref{eq:numax} and \ref{eq:largesep} with the mass, radius and  effective temperature obtained from the evolutionary calculation.
The resulting values for $\chi^2$ are listed in Table\,\ref{tab:chi2} and can be seen in Fig.\,\ref{fig:opt}. The mass estimate is $M=(3.9\pm0.1)M_{\odot}$ and the other resulting model parameters are listed in Table\,\ref{tab:chi2}. A given parameter can be both a constraint with an uncertainty (thus having to be in a given range), and an output. 

\begin{table}
\center
\caption{Characteristics of the models presenting the lowest $\chi^2$ value in Eq.\,\ref{eq:chi2} for a given physics and chemical composition and when using a varying number of constraints (see text).}
\label{tab:chi2}
\begin{tabular}{c|c|c|c}
\hline
Constraints     & $\langle\Delta\nu\rangle$ & $\langle\Delta\nu\rangle$, $\nu_{\rm max}$ & $\langle\Delta\nu\rangle$, $\nu_{\rm max}$\\
used		       & $\nu_{\rm max}$ & $T_{\rm eff}$ & $T_{\rm eff}$, $\log L/L_\odot$ \\
\hline
$\chi^2$ & 0.008 & 0.36 & $\le$ 1.1 \\
$M/M_{\odot}$ & 3.6 & 3.3 & 3.8 -- 4.0 \\
$R/R_{\odot}$ & 28.3 & 27.5 & 29.4 -- 30.9 \\
\hline
\end{tabular}
\end{table}

When the luminosity is taken into account, the different $\chi^2$ values are significantly higher and form a broader shape in Fig.\ref{fig:opt}: low values of $\chi^2$ can be reached for several models.
The photometric luminosity constraint is less robust because of the uncertainty on the parallax of the star. However, it contributes additional information on the radius and thus should be taken into account. The consequence is a larger uncertainty on the mass determination as the lower values of $\chi^2$ can be reached for models of different masses. 

\subsection{Influence of the physical description on the modelling}
\label{sec:model2}

A change in the assumed physical   description is likely to modify the  stellar  characterics of the ``best'' models determined in the previous section. We illustrate this issue by assessing the effect of changing the values of the free parameters related to convective motions, $\alpha_{\rm CGM}$ and $\alpha_{\rm ov}$, as well as the metallicity $Z$.

\begin{figure}
\centering
\includegraphics[trim= 100 150 100 100,width=6cm,angle=0]{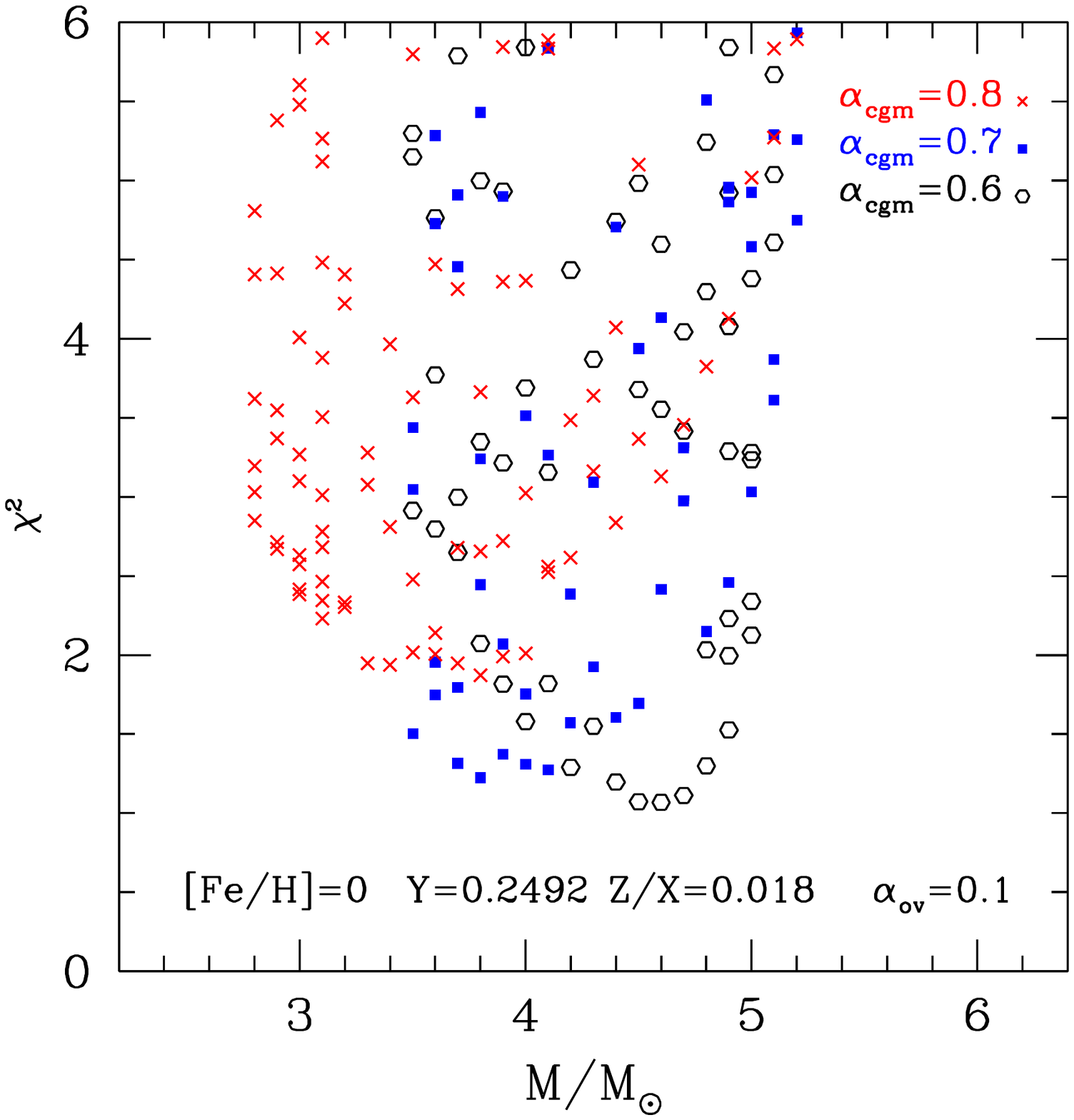}
\includegraphics[trim= 100 150 100 100,width=6cm,angle=0]{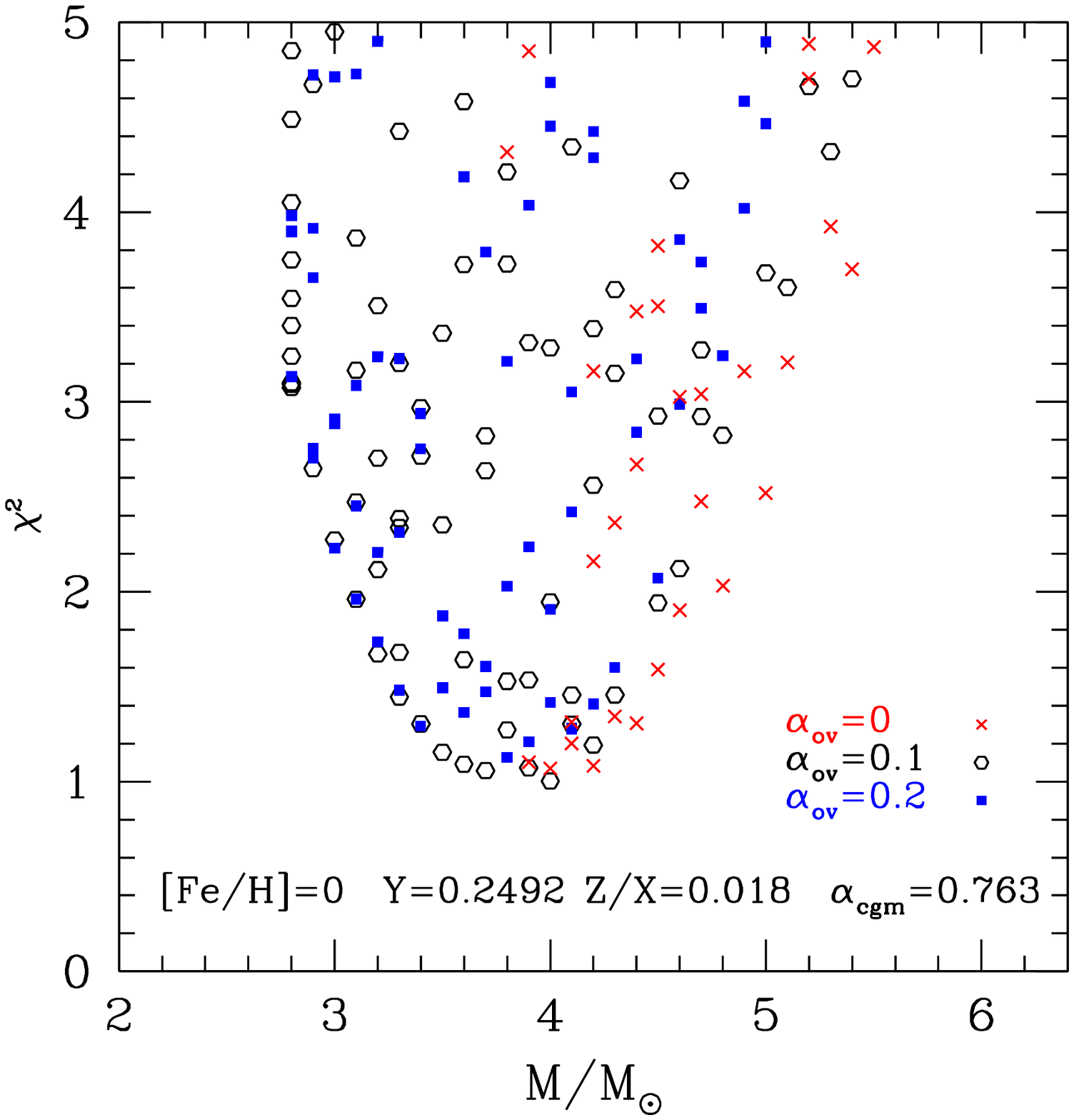}
\includegraphics[trim= 100 150 100 100,width=6cm,angle=0]{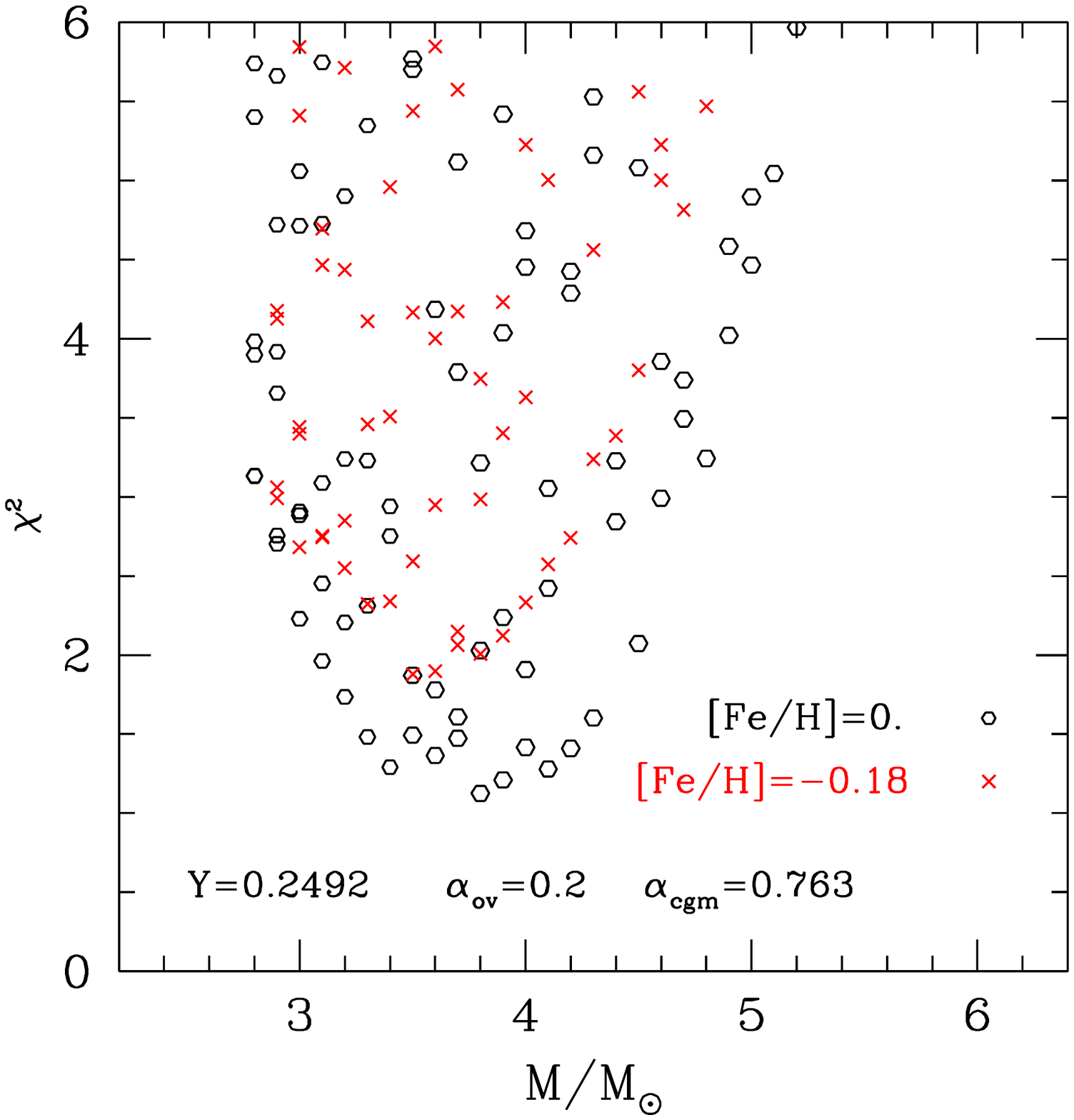}
\caption{Computed values of $\chi^2$ (Eq.\,\ref{eq:chi2}) versus the mass, when one parameter of the modelling is varied while the others are kept constant: $\alpha_{\rm CGM}$ (top, $M/M_{\odot}=4.6\pm0.2$), $\alpha_{\rm ov}$ (middle, $M/M_{\odot}=3.9\pm0.3$) and ${\rm [Fe/H]}$ (bottom, $M/M_{\odot}=3.8\pm0.2$). The final estimate is $3.6M_{\odot}<M<4.8M_{\odot}$.}
\label{fig:compchi2}
\end{figure}

An illustration of the influence of varying  $\alpha_{\rm CGM}$ (with all other parameters kept constant) on evolution tracks in the H-R diagram is shown in Fig.\,\ref{fig:acgm}. The models satisfying the observational constraints in the H-R diagrams shift to lower luminosity and temperature with $\alpha_{\rm CGM}$ increasing. In order to take into account the influence of other parameters,
model grids with $\alpha_{\rm ov}=0$ and $\alpha_{\rm ov}=0.2$ were computed, everything else being kept the same as for $\alpha_{\rm ov}=0.1$, as in Sect.\,\ref{sec:model1}. Using the same approach as above, $\chi^2$ (for all four constraints) are computed and displayed in Fig.\,\ref{fig:compchi2} where the effect of the variation of $\alpha_{\rm CGM}$ on mass determination appears clearly: a higher $\alpha_{\rm CGM}$ leading to a lower mass, the variation being as high as 1.0\,$M_{\odot}$ for a variation of 0.2 in $\alpha_{\rm CGM}$. The smallest value of $\alpha_{\rm CGM}$=0.6 provides the lowest values of $\chi^2$.
Proceeding in the same manner for variations of $\alpha_{\rm ov}$, mass variations are smaller but still of the order of 0.2\,$M_{\odot}$. The different values of $\alpha_{\rm ov}$ lead to similar values of $\chi^2$.
The metallicity  also has an influence on the mass determination: varying  the solar metallicity to  ${\rm [Fe/H]}=-0.18$ the difference in mass reaches approximately 0.2\,$M_{\odot}$. The lowest $\chi^2$ values are clearly in favour of the solar metallicity, but even lower values could be obtained for intermediate values of the metallicity.

The different stellar characteristics, yielded by estimating the lowest $\chi^2$ values in a grid of models for the different parameters considered here, are  summarized in Table\,\ref{tab:sensi}. These values do not aim at being precise (the precise outputs of the optimum model are given in  Sect.\,\ref{sec:optmodel}) but at showing the approximate range of the model parameters.

\begin{table}
\center
\caption{Sensitivity of the stellar characteristics to the model parameters. The  estimates presented correspond to the models giving the lowest $\chi^2$.}
\label{tab:sensi}
\begin{tabular}{c|ccc}
\hline
$\alpha_{\rm ov}$ & 0 & 0.1 & 0.2 \\
\hline
$M/M_{\odot}$ & 3.9 -- 4.2 & 3.6 -- 4.0 & 3.6 -- 4.0 \\
$R/R_{\odot}$ & 29.5 -- 30.5 & 28 -- 30 & 27.5 -- 30 \\
\hline
\end{tabular}
\vskip2mm
\begin{tabular}{c|ccc}
\hline
$\alpha_{\rm CGM}$ & 0.6 & 0.7 & 0.8 \\
\hline
$M/M_{\odot}$ & 4.4 -- 4.8 & 3.6 -- 4.1 & 3.3 -- 4.0 \\
$R/R_{\odot}$ & 31 -- 32 & 29 -- 30 & 27 -- 29 \\
\hline
\end{tabular}
\vskip2mm
\begin{tabular}{c|cc}
\hline
 ${\rm [Fe/H]}$ & 0 & -0.18 \\
 \hline
 $M/M_{\odot}$ & 3.6 -- 4.0 & 3.5 -- 3.7 \\
$R/R_{\odot}$ & 27.5 -- 31 &  27 -- 28\\
\hline
\end{tabular}
\caption{These estimates are drawn from the results of Fig.\,\ref{fig:compchi2} for the mass and from similar results not shown for the radius. They do not aim at a quantitative comparison but at an overall view only.}
\end{table}

\subsection{Optimal modelling}
\label{sec:optmodel}

As shown in the previous section, the search of an optimal model should 
allow all of the parameters related to the physical description to vary. However, this can suffer from limitations mentioned below. 
The determination of the best model used here is performed with the method described in \citet{Miglio05}: $\chi^2$ is determined from a minimisation process using a Levenberg--Marquardt algorithm coupled to the CESAM algorithm in order to converge to the parameter values that provide the lowest $\chi^2$. The searched models are restricted to the ascending branch of red giants and do not consider the further evolution of the star (He-core burning, this  will be discussed in Sect.\,\ref{sec:age}).

However, the application of this method cannot avoid some limitations. First, many parameters are needed to fully describe a model whereas there are in comparison only a few independent observational constraints. If all parameters are  free, the result of the minimisation is not reliable as it depends on the initial values of the parameters. This led us to run the minimisation process for a grid of fixed values of a given parameter (see below). Second, there is some cross-talk between the model parameters (for example between the mass and $Y$, and also between $\alpha_{\rm CGM}$ and $Y$). The consequence is a degeneracy of the solution.

\begin{figure}
\centering
\includegraphics[width=7cm]{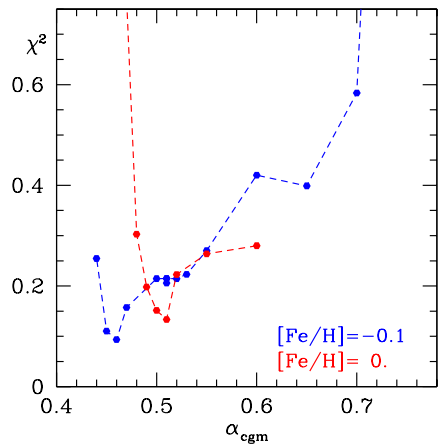}
\caption{Variation of $\chi^2$ in the minimisation process for different fixed values of $\alpha_{\rm CGM}$ (and $\alpha_{\rm ov}=0.1$, ${\rm [Fe/H]}=0$ (in blue) and -0.1 (in red) fixed, whereas the mass, age and initial helium abundance are free parameters).}
\label{fig:chi2alphaCGM}
\end{figure}

\begin{figure}
\centering
\includegraphics[width=7cm]{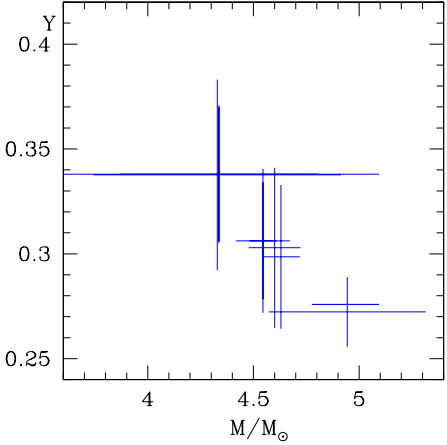}
\caption{Values of $Y$ versus $M$ for models obtained with the minimisation processes (with fixed values of $\alpha_{\rm CGM}$) and closest to the constraints ($\chi^2<0.3$): a clear anti-correlation appears.}
\label{fig:YM}
\end{figure}

A first set of minimisation processes was run with different values of ${\rm [Fe/H]}$. It appears that extreme values of the metallicity within the observational range lead to values of $Y$ either lower than the primordial He abundance (${\rm [Fe/H]}=-0.18$ yields $Y=0.24$), or too high (${\rm [Fe/H]}=0$ yields $Y=0.30$, a value that corresponds to those of the youngest open cluster stars and therefore lies in the upper part of the reasonable range). Consequently, the minimisation will be then processed using a fixed, intermediate  value of the metallicity: ${\rm [Fe/H]}=-0.1$.\\
Then, the minimisation process was run with fixed values of the metallicity, $Y$, $\alpha_{\rm CGM}$ and $\alpha_{\rm ov}$ which could take three values: 0, 0.1 and 0.2. The best values of $\chi^2$ were obtained for $\alpha_{\rm ov}=0.1$. The next parameter to be investigated was $\alpha_{\rm CGM}$. A series of minimisation processes was run for decreasing values of $\alpha_{\rm CGM}$ (from 0.75 to 0.44, but fixed for each minimisation process run) whereas $M$, $Y$ and the core temperature (from which the age is derived) are free parameters. A clear trend appears when selecting the models giving the lowest values of $\chi^2$ (smaller than 0.3): they correspond to low values of  $\alpha_{\rm CGM}$ (smaller than 0.6, see Fig.\,\ref{fig:chi2alphaCGM}). In addition, when selecting these models ($\chi^2<0.3$), a clear anti-correlation appears between their $M$ and $Y$ values (see Fig.\,\ref{fig:YM}).\\
A similar relation is observed between $\alpha_{\rm CGM}$ and $Z$.
For the optimum model, $\alpha_{\rm CGM}\simeq0.45\pm0.15$.
It is obtained for a fixed metallicity: ${\rm [Fe/H]}=-0.1$. If the metallicity is fixed at ${\rm [Fe/H]}=0$, then $\alpha_{\rm CGM}\simeq0.51\pm0.15$ (see fig.\,\ref{fig:chi2alphaCGM}). Both values are substantially lower than the solar value ($\alpha_{\rm CGM,\odot}\simeq0.76$) but are fully compatible with results from 3D numerical simulations of convection in red giants ($\alpha_{\rm CGM,\odot}\simeq0.6$, Samadi et al. -- in prep.)

The characteristics of the optimal model (lowest $\chi^2$) are listed in Table\,\ref{tab:optim}, but one should not forget that this model corresponds to given values of metallicity and $\alpha_{\rm ov}$.

$\langle\Delta\nu\rangle$ has been computed from the mode frequencies given by the optimum model, and an expected $\nu_{\rm max}$ is then computed using the scaling relation between these two parameters. As seen from Table\,\ref{tab:optim}, the computed $\nu_{\rm max}$ is in agreement with the observed value.

A last remark on this optimal model: it was found with a metallicity in the upper part of the interval given by observations. As mentioned in Sect.\,\ref{sec:spectro}, the parameters $T_{\rm eff}$ and ${\rm [Fe/H]}$ were determined without strong constraints on $\log g$. Using the values of $M$ and $R$ from Table\,\ref{tab:sensi}, it was then possible to iteratively improve the atmospheric parameters using $\log g$ from the seismic modelling, which is close to $2.09\pm0.02$. For a fixed value of $\log g$, we found $T_{\rm eff}=4700\pm180$\,K and ${\rm [Fe/H]} = -0.05 \pm 0.11$. This higher observed metallicity obtained a posteriori is in agreement with seismic modelling.

\begin{table}
\center
\caption{Characteristics of the optimal model (giving $\chi^2=9.35\cdot10^{-2}$). Uncertainties are listed in Table\,\ref{tab:recap}.}
\label{tab:optim}
\begin{tabular}{cc}
\hline
\multicolumn{2}{c}{Constraints}\\
% & \\
$\log(L/L_{\odot})$ & 2.70 $\pm$ 0.15\\
$\log(T_{\rm eff})$ & 3.67 $\pm$ 0.02 \\
$\nu_{\rm max}$ ($\mu$Hz) & 15.0 $\pm$ 1.0 \\
$\Delta\nu$ ($\mu$Hz) & 1.7 $\pm$ 0.1\\
 \hline
\multicolumn{2}{c}{Fixed parameters}\\
% & \\
${\rm [Fe/H]}$   & -0.1 \\
$\alpha_{\rm ov}$ & 0.1 \\
$\alpha_{\rm CGM}$ & 0.46 \\
\hline
\multicolumn{2}{c}{Fitted parameters}\\
% & \\
$M/M_{\odot}$ &  4.63 $\pm$ 0.09 \\ 
$Y$ & 0.298 $\pm$ 0.034\\ 
$T_{\rm core}$ (MK) & 80.3 $\pm$ 14 \\
 \hline
\multicolumn{2}{c}{Outputs of the model}\\
% & \\
Age (Myr) & 105.5  \\
$\log(L/L_{\odot})$ & 2.66 \\
$\log(T_{\rm eff})$ & 3.67 \\
$\nu_{\rm max}$ ($\mu$Hz) & 15.0 \\
$\Delta\nu$ ($\mu$Hz) & 1.69 \\
$R/R_{\odot}$ & 32.27 \\
\hline
\end{tabular}
\end{table}

\subsection{Discussion of the accuracy of  HR\,2582 (HD\,50890) stellar parameter determination}
\label{sec:deltamass}

As discussed in the previous sections, there are several ways to estimate the mass of a star, and this determination can rely on different observational constraints. We have first used model grids. The corresponding evolution tracks in the H--R diagram can be compared to the observed luminosity and temperature and their error bars. Then, seismic constraints ($\nu_{\rm max}$ and $\Delta \nu$) can also be taken into account to select the adequate evolution tracks in the H--R diagram. This selection can be quantified by looking at the sum of squared differences $\chi^2$ between observed and computed parameters from a grid of models (see Sect.\,\ref{sec:model1}). This could be performed with a grid of models computed for a given ``physics description'' (parameters describing the overshoot, the convection and also a given metallicity and helium abundance) or allowing this physics to vary (see Sect.\,\ref{sec:model2}). Finally, an optimal modelling can be performed, allowing the physics parameters to vary during the process of the minimisation of $\chi^2$ (see Sect.\,\ref{sec:optmodel}). To be complete, we recall that the large separation as an output from modelling can result from the use of the scaling relation (Eq.\,\ref{eq:largesep}) and the modeled effective temperature, or directly from the computed modeled frequencies.

We could have also used the seismic constraints only, together with the observed temperature and
combining the scaling relations of Eq.\,\ref{eq:numax} and \ref{eq:largesep}. Then, one obtains the following uncertainties on the mass determination:
\begin{equation}
\frac{\sigma(M)}{M} = 3 \frac{\sigma(\nu_{\rm max})}{\nu_{\rm max}}+4\frac{\sigma (\Delta\nu)}{\Delta \nu}+{3 \over 2 }\frac{\sigma(T_{\rm eff})}{T_{\rm eff}}
\end{equation}
and similarly for the radius:
\begin{equation}
\frac{\sigma(R)}{R} = \frac{\sigma(\nu_{\rm max})}{\nu_{\rm max}}+2\frac{\sigma(\Delta \nu)}{\Delta \nu}+{1 \over 2 }\frac{\sigma(T_{\rm eff})}{T_{\rm eff}}
\end{equation}
However, these scaling relations are  empirical  and hence suffer from some uncertainty. The corresponding estimates are given in Table\,\ref{tab:recap}, together with all of the other resulting mass estimates and their uncertainties. One can draw some conclusions and some limitations from this table. As expected, the use of the seismic constraints together with the temperature and luminosity  substantially improves the mass determination. However, seismic constraints alone (using the scaling relations) do not yield a precise estimate. Such a precise estimated is given by the use of a grid of models, which can however lead to a largely under-estimated uncertainty if the model physics is fixed.

\begin{table*}
\center
\caption{Summary of mass estimation and uncertainty. $\Delta \nu$ represents the large separation derived from Eq.\ref{eq:largesep}, $\langle\Delta \nu\rangle$ represents the large separation computed from the modeled frequencies. $T_{\rm eff}^{\rm obs}$ and $T_{\rm eff}^{\rm mod}$ are respectively the observed and modeled temperatures.}
\label{tab:recap}
\begin{tabular}{cccccc}
\hline\\[-1ex]
 &  & $M$ & $\sigma(M)$ & $\sigma(M)/M$ (\%) & Mass interval ($M_{\odot}$)\\[1ex]
\raisebox{2ex}{Method} & \raisebox{2ex}{Parameters used} & $R$ & $\sigma(R)$ & $\sigma(R)/R$ (\%) & Radius interval ($R_{\odot}$)\\[1ex]
\hline
\hline\\[-1ex]
Model grids and  &  & 3.5 &2 & 57 & $1.5< M <5.5$ \\[-1ex]
H-R diagram        &  \raisebox{2ex}{$L^{\rm obs}$, $T_{\rm eff}^{\rm obs}$}  &  34 & $\sim$8 & 24 & $26 < R < 42$ \\[1ex]
\hline\\[-1ex]
Scaling relations & & 3.6 & 1.8 & 50 & $1.8<M<5$ \\[-1ex]
(using Eq.\ref{eq:numax} and \ref{eq:largesep} with $T_{\rm eff}^{\rm obs}$) & \raisebox{2ex}{$\Delta \nu$, $\nu_{\rm max}$} & 28.8 & 6 & 21 & $22.6<R<34.6$ \\[1ex]
\hline\\[-1ex]
Model grid with fixed input parameters& & 3.9 & 0.1 & 2.5 &$3.8<M<4.0$ \\[-1ex]
(using Eq.\ref{eq:numax} and \ref{eq:largesep}
 with $T_{\rm eff}^{\rm mod}$, $T_{\rm eff}^{\rm mod}$ compatible with $T_{\rm eff}^{\rm obs}$)  & \raisebox{2ex}{$L^{\rm obs}$, $T_{\rm eff}^{\rm obs}$, $\Delta \nu$,
 $\nu_{\rm max}$}   & 29.2 & 0.3 & 1.0 & $29.0<R<29.5$ \\[1ex]
\hline\\[-1ex]
Model grid with varying input parameters  & & 4.2 & 0.6 & 14 & $3.6<M<4.8$ \\[-1ex]
(using Eq.\ref{eq:numax} and \ref{eq:largesep} with $T_{\rm eff}^{\rm mod}$, $T_{\rm eff}^{\rm mod}$ compatible with $T_{\rm eff}^{\rm obs}$) &
\raisebox{2ex}{$L^{\rm obs}$, $T_{\rm eff}^{\rm obs}$, $\Delta \nu$, $\nu_{\rm max}$} & 30.0 & 1.4 &
4.7 & $28.5<R<31.3$ \\[1ex]
\hline\\[-0.5ex]
Optimal modelling & &4.2 &0.9 & 21.5 & $3.3<M<5.3$ \\[-1ex]
(using Eq.\ref{eq:largesep} with $T_{\rm eff}^{\rm mod}$, $T_{\rm eff}^{\rm mod}$ compatible with $T_{\rm eff}^{\rm obs}$) & 
\raisebox{2ex}{$L^{\rm obs}$, $T_{\rm eff}^{\rm obs}$, $\Delta \nu$, $\nu_{\rm max}$} & 29.9 & 1.9 & 6.3 & $28.0<R<31.8$ \\[1ex]
\hline
\end{tabular}
\end{table*}

\subsection{Echelle diagram}
\label{sec:diagech}

The \'echelle-diagram built  with the frequencies  of the  optimal model is compared to the  observed one in Fig.\,\ref{fig:diagech}. The theoretical  \'echelle-diagram (computed without corrections of near-surface effects) matches  the observed one well within the observational error bars. Note that the folding value for the model, 1.715\,$\mu$Hz, is only slightly different from that of the observations (1.7\,$\mu$Hz). This value, usually associated with the mean large separation is different from the value 1.7\,$\mu$Hz given by Eq.\,\ref{eq:largesep} which is the relation used in the minimisation process. It nevertheless falls within  the uncertainties associated with this relation. The present modelling also confirms the observed values of radial order ($5 \le n \le 12$).

\begin{figure}
\centering
\includegraphics[trim= 100 150 100 100,width=7cm]{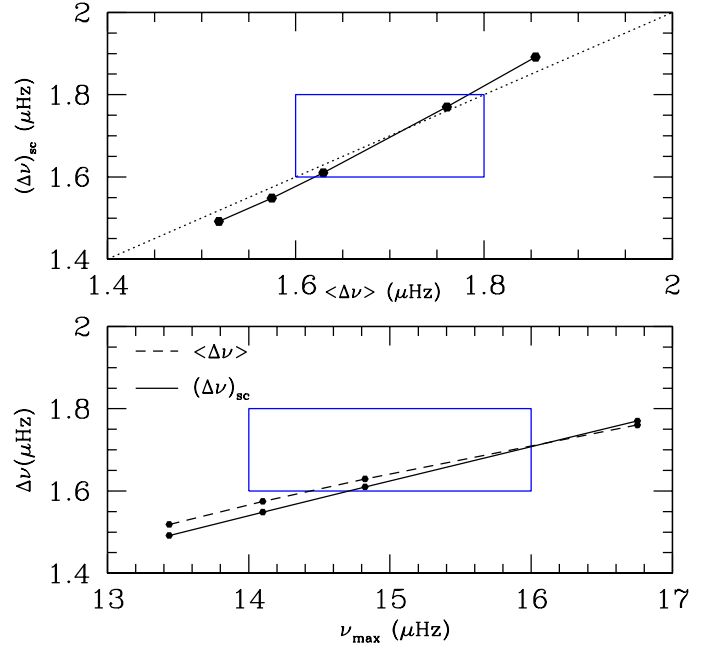}
\caption{Upper panel, solid line: $\Delta\nu_{\rm sc}$, computed from the scaling relation (Eq.\,\ref{eq:largesep}) and the modeled mass and $T_{\rm eff}$ versus $\Delta\nu$ computed from the actual frequencies given by the optimal model (the blue frame indicates the observations); lower panel: $\Delta\nu_{\rm sc}$ and $\Delta\nu$ versus the computed $\nu_{\rm max}$ from Eq.\,\ref{eq:numax} for a series of models satisfying  the observations represented as the blue frame. This validates the use of the scaling relations in the computations.}
\label{fig:meandeltanu}
\end{figure}

We also checked that the estimation of $\Delta \nu$ from a scaling relation and the modeled mass (Eq.\,\ref{eq:largesep}) was correct by comparing with the frequencies derived from the model (see Fig.\,\ref{fig:meandeltanu}).

\section{Evolutionary stage ambiguity}
\label{sec:age}

\begin{figure}
\centering
\includegraphics[trim= 100 150 100 100,width=7cm]{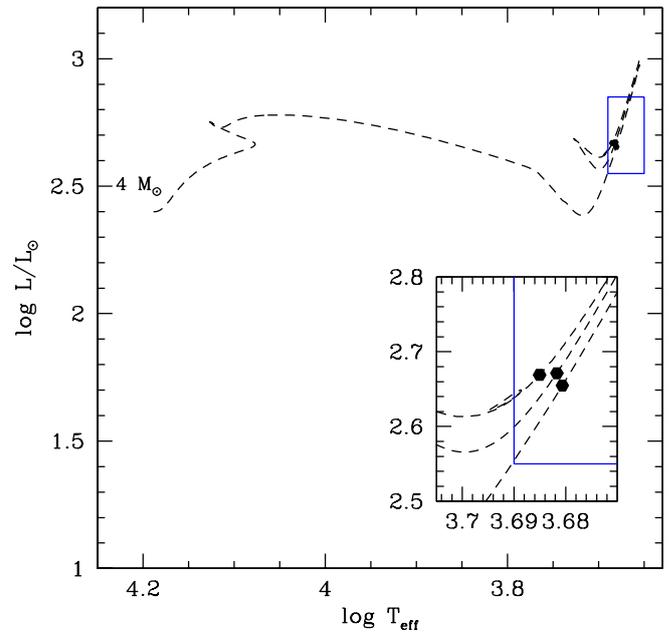}
\caption{Evolutionary track  for a 4\,$M_\odot$ model showing that it returns to the same location of the H-R diagram at the beginning of the ascending branch (lowest track) and during He-core burning (descending -- intermediate track -- and second ascending (higher track) branches). Selected models discussed in the text are plotted with  solid black dots.}
\label{fig:HRage}
\end{figure}

\begin{figure}
\centering
\includegraphics[trim= 100 150 0 100,width=7cm]{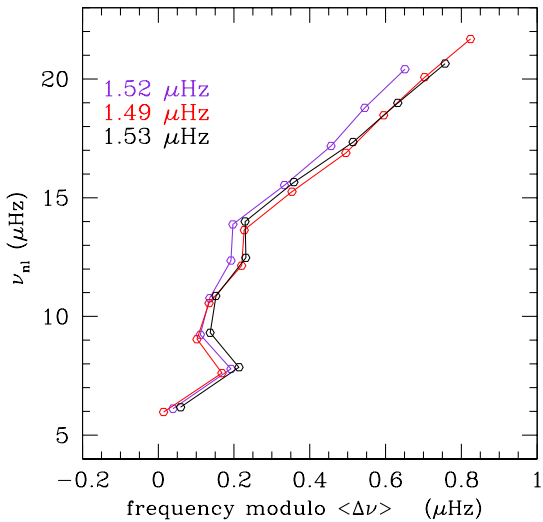}
\caption{Echelle diagram for the 4\,$M_\odot$ models of Fig.\ref{fig:HRage}. $\langle\Delta\nu\rangle$ is taken to be 1.53, 1.49 and 1.52\,$\mu$Hz (for the observed range: $5 \le n \le 12$) for the three stages of evolution:  the youngest (violet), intermediate (red) and the oldest (black) ones. The differences between the three cases are too small to be detected with the present data set.}
\label{fig:diagech2}
\end{figure}

As mentioned earlier, the location of HR\,2582 (HD\,50890)  in the H-R diagram can correspond to several evolutionary stages: either on the H-shell burning on the first ascending branch or He-core burning on the descending or second ascending branch (see Fig.\,\ref{fig:HRage}). The respective ages for these three evolutionary stage are 157, 163 and 180 Myr. The corresponding helium abundance in the core are $Y_{\rm c}=0.983$ (no helium fusion yet), 0.768 and 0 (helium fusion phase ended).
At a given location the star has similar mass and radius and  
the large separation $\Delta\nu$  is unable to discriminate between the possible evolutionary stages, as discussed by \citet{Mazumdar09}.
To emphasise this problem in the present case, we have considered a 4\,$M_{\odot}$ star computed assuming $\alpha_{\rm CGM}=0.7$, $\alpha_{\rm ov}=0.15$, ${\rm [Fe/H]}=0$, $Z=0.0172$, $Y=0.28$ using the solar mixture of \citet{Grevesse93}.
The values of $\langle\Delta\nu\rangle$ (computed as the mean separation of modes with $5 \le n \le 12$) for the three stages of evolution are 1.53, 1.49 and 1.52\,$\mu$Hz, falling in a quite narrow range, smaller than the frequency resolution of the present data set ($\delta\nu=0.21\mu$Hz).
The three models do not have  exactly the same radius, hence not exactly the same mean large separation. This explains why the  $\ell=0$ ridges in the \'echelle diagram shown in Fig.\,\ref{fig:diagech2} are slightly shifted from each other. If one scales the offset with the respective mean large separation  for the three  cases, the variations of the three  ridges  with the frequency nearly coincide.
This confirms the expectation that the variations of the ridges with the
frequency arise from the properties of the surface layer which are similar for the three models (see Fig.\,\ref{fig:diagech2}). 

If non-radial mixed modes were present, and if the time series were long enough to provide a sufficient frequency resolution to resolve period spacings of these mixed modes, the age ambiguity could be removed thanks to the age signature in mixed-mode spacings as shown by \citet{Bedding11} and \citet{Mosser11b}.
However, for a star with such a low $\langle\Delta\nu\rangle$ value, this requires an observing time longer than 2 years.

\section{Conclusions}
\label{sec:conclu}

HR\,2582 (HD\,50890) is a luminous giant that shows a number of peaks in the  power spectrum at low frequency, centred around the frequency $\nu_{\rm max}=15\,\mu$Hz, with a mean frequency separation $\langle\Delta\nu\rangle=1.7\,\mu$Hz. This separation can be expected from scaling relations (Eq.\,\ref{eq:numax} and \ref{eq:largesep}) and implies that only radial modes are  clearly observed.
\citet{deRidder09} and\cite{Bedding10} clearly showed the presence of non-radial modes in red giants. Their absence in our observations could be caused by a too short observation sequence or, since our results for HR\,2582 (HD\,50890) lead to an estimated mass around 4\,$M_{\odot}$, the internal structure (namely the ratio of the density of the core to the mean density) could be different and lead to different amplitudes for non-radial modes in high-mass stars.

%In addition to the observational constraints on the luminosity and effective temperature of HR\,2582 (HD\,50890), the use of the seismic constraints $\nu_{\rm max}$ and $\Delta\nu$ allows the restriction of the range of possible values for the model parameters (mass, radius, age...) used to describe the observed star. The seismic constraints definitely exclude the higher luminosity and the lower temperature ranges, allowing a more constrained modelling. However, were the estimated effective temperature be underestimated,
%%by 500\,K,
%the actual luminosity could then certainly be higher.

The modelling of the star is shown to be sensitive to the physical description used (parameters describing  convection,  overshoot, etc...) and other constraints such as the metallicity for example. 
%The modelling is subject to the physical description used (parameters describing  convection,  overshoot, etc...) and other constraints such as the metallicity for example. 
%Thus, an efficient modelling of the star must allow for the description of the physics that includes its uncertainties, and the resulting uncertainties on the output must be estimated.
For example, the value of the estimated mass can vary by about 1\,$M_{\odot}$ for different parametrization of the convection. Only some aspects of the physics involved in stellar modelling were varied in the present work. Thus, the uncertainties derived from modelling are possibly under-estimated: for example, the influence of microscopic diffusion was not taken into account here.

An optimal model was found with a mass of $(4.15 \pm 0.85) M_{\odot}$.
%, confirming that this red giant does not belong to the so-called red clump (low mass He-burning stars).
%This star is thus an interesting case in order to complete the sampling of the modelling of the red giant population.
Supposing the star is on the ascending red giant branch, its age was estimated from this optimal model to be 155 Myr.
Another interesting result is the low value of  $\alpha_{\rm CGM}$, lower than both the solar value and results obtained from simulations, indicating that the modelling of red giants could bring in return particular inputs on the description of convection.\\

The \'echelle-diagram built from the optimal model is globally in agreement with the observed one (Fig.\,\ref{fig:diagech}).
%We checked that the estimation of $\Delta \nu$ from a scaling relation and the modeled mass (Eq.\,\ref{eq:largesep}) was correct by comparing with the frequencies derived from the model (see Fig.\,\ref{fig:meandeltanu}).
The small differences between modeled and observed \'echelle-diagrams contain potentially more information about the internal structure of the star and will be investigated in a forthcoming work. 
For example, microscopic diffusion is known to lead to  strong  element depletion for such massive stars. Then, turbulent mixing should be present to compensate for such a non-observed high element depletion. It is possible that seismology of high mass stars such as HR\,2582 (HD\,50890) can bring some constraints on this issue (to be investigated in a future work).

\begin{acknowledgements}
TM acknowledges financial support from Belspo for contract PRODEX GAIA-DPAC. SH acknowledges financial support from the Netherlands Organisation for Scientific Research (NWO). TK acknowledges the support of the FWO-Flanders under project O6260 - G.0728.11. This work used data from GAUDI, the data archive and access system of the ground-based asteroseismology programme of the COROT mission. The GAUDI system is maintained at LAEFF. LAEFF is part of the Space Science Division of INTA. FB wish to thank L.M.R. Lock for reading and improving the manuscript.
\end{acknowledgements}

\bibliographystyle{aa} % style aa.bst
\bibliography{biblio} % your references Yourfile.bib

\end{document}